\documentclass[pra,amsmath,aps,10pt,letterpaper,tightenlines,showpacs]{revtex4} 
\usepackage{bm}
\usepackage{stmaryrd}
\usepackage{amsmath}
\usepackage{amssymb}
\usepackage{graphicx}
\usepackage{textcomp}
\usepackage{calrsfs}
\usepackage{yfonts}
\usepackage[breaklinks=true,colorlinks=true,linkcolor=blue,urlcolor=blue,citecolor=blue]{hyperref}

\begin{document}
\title{Cooling of a Nanomechanical Resonator in Presence of a Single Diatomic Molecule}
\author{M. Eghbali-Arani$^{1}$}
\author{Sh. Barzanjeh$^{2}$}
\author{H. Yavari$^{1}$}
\email{hs-yavary@yahoo.com}
\author{M. A. Shahzamanian$^{1}$}
\affiliation{
$^1$Department of Physics, Faculty of Science, University of Isfahan, Hezar Jerib, 81746-73441, Isfahan, Iran \\
$^2$Institute for Quantum Information, RWTH Aachen University, 52056 Aachen, Germany}

\begin{abstract}
We propose a theoretical scheme for coupling a nanomechanical resonator to a single diatomic molecule via microwave cavity mode of a driven $ LC $ resonator. We describe the diatomic molecule by a Morse potential and find the corresponding equations of motion of the hybrid system by using Fokker-Planck formalism. Analytical expressions for the effective frequency and the effective damping of the nanomechanical resonator are obtained. We analyze the ground state cooling of the nanomechanical resonator in presence of the diatomic molecule. The results confirm that presence of the molecule improves the cooling process of the mechanical resonator. Finally, the effect of molecule's parameters on the cooling mechanism is studied.
\end{abstract}

\maketitle



\section{Introduction}
%
%
In optomechanical cavity, radiation pressure acts on an oscillating mirror which induces an interaction between the mechanical system and the optical field.
The coupling of mechanical and optical degrees of freedom via radiation pressure has been employed for a wide range of applications, such as the cavity cooling of microlevers and nanomechanical resonators to their quantum mechanical ground states~\cite{Aspelmeyer,Chen, Cleland,Kleckner2006,Poggio2007,Teufel2008,Arcizet2006,Teufel2011}. 

Activity in this field started with experimental observations of optomechanical cooling first using feedback~\cite{PhysRevLett.83.3174}, and later using an intrinsic effect~\cite{gigan2006self}. Aside from optomechanical cooling, electronic cooling was also studied. For instance, schemes have been proposed  
to replace the optical cavity by radio frequency circuits~\cite{brown2007passive} or 
one-dimensional transmission line resonators~\cite{xue2007cooling,zhang2009cooling}.

Very recently, various schemes have been also proposed in order to couple mechanical resonator to other systems~\cite{Xiang2013} including single atoms~\cite{Hammerer2009,wallquist2010single,chang2009triple,breyer2012light,barzanjeh2011steady}, 
atomic ensembles~\cite{ian2008cavity,schleier2011optomechanical,Genes2008}, ions~\cite{Tian2004}, molecules~\cite{bhattacharya2010optomechanical,singh2008coupling}, 
and electrons~\cite{PhysRevA.75.032348}. In this direction, as is shown in Ref.~\cite{ian2008cavity}, the presence of an atomic ensemble effectively enhances the optomechanical coupling rate.
However, a theoretical description of the motion of mechanical resonator based on the capacitive coupling of the resonator with 
a superconducting coplanar waveguide, was discussed in Ref.~\cite{PhysRevA.76.042336}, 
which focused on studying the entanglement between mechanical resonator and transmission line resonator without considering the cooling of mechanical resonators. 
Since the quantized electric field of a resonator circuit can be easily coupled to ions~\cite{kielpinski2012quantum} and atoms~ \cite{Daniilidis2013}, then 
the direct coupling between the electrical circuits and mechanical resonator also utilizes a new way for coupling mechanical resonators either to two-level systems 
or ions~\cite{Xiang2013}. It also paves the way for coupling the  mechanical resonators to the dipole moment of the diatomic molecules. The diatomic molecules usually can  be described by a nonlinear Morse potential. The interaction between a weakly nonlinear Morse oscillator to quantized intracavity field was also studied in Refs.~\cite{Gan1990,Gan1991}. The authors used the Jaynes-Cummings like interaction between molecule and electric field in which the dipole of the molecule interacts with electric field of the cavity mode.

Motivated by the 
aforementioned studies, the basic study here is a Morse oscillator, the simple diatomic molecular system with 
non-equidistant multi-level states, coupled to a nanomechanical resonator via a microwave cavity mode of a driven superconducting 
LC resonator. Both the dynamics of the nanomechanical oscillator and the properties of microwave field are modified through this interaction.  
The purpose of this paper is to investigate the effect of presence of the diatomic molecule on the cooling of the nanomechanical resonator. The Morse oscillator is assumed to be weakly non-linear, and attains an equilibrium with the driving field through the effect of radiation damping only. We omit all other relaxation processes and neglect the wave mixing effects from the present analysis.
We derive an exact second-order Fokker-Planck equation for the hybrid system. The Fokker-Planck 
equation can be converted into an equivalent set of first-order stochastic differential equations, which can be used for studying the cooling of the mechanical resonator.

The paper is structured as follows: In Sec. II, first we
present a hybrid system composed of a microwave cavity mode, a nanomechanical resonator, and a diatomic molecule then the Hamiltonian of the hybrid configuration is found. In Sec. III, the stochastic equations corresponding to 
the Fokker-Planck equation is obtained by It\^{o}'s rule and the dynamics of the systems is discussed. In Sec. IV, we investigate the effective frequency and
the effective damping parameter of the nanomechanical oscillator. In Sec. V, the ground state cooling of the nanomechanical resonator is discussed. Finally, a summary and concluding remarks are given in
Sec. VI.

%
\section{SYSTEM MODEL AND HAMILTONIAN}
In this section, first we shortly discuss about the Hamiltonian of a diatomic molecule, then we find the Hamiltonian of a hybrid system in which a nanomechanical resonator couples to a diatomic molecule via microwave cavity mode.
 \subsection{Morse potential}
It is known that the interaction between atoms in a diatomic molecule can be described by anharmonic Morse potential~\cite{dong2003controllability,angelova2008generalized}. The Morse potential can be expressed as~\cite{dong2003controllability}
\begin{eqnarray}\label{Morse}
V(r)=D_e (1-e^{-a(r-r_e)})^2,
\end{eqnarray}
where $r$ is the inter-nuclear distance between the atoms, $r_e$ is the equilibrium bond distance, $D_e$ 
is the well depth (defined relative to the dissociated atoms), and $a=\omega_e\sqrt{\mu/2D_e}$ is related with the range of the potential which identifies by
 reduced mass $\mu$ and fundamental vibrational frequency $\omega_e$. The Schr\"{o}dinger equation for 
the Morse potential is exactly solvable, giving the vibrational eigenvalues
\begin{eqnarray}\label{Morse}
E_{\nu}=\omega_e (\nu+1/2)-\frac{\omega_e^2}{4D_e}(\nu+1/2)^2,
\end{eqnarray}
 where $\nu=0,1,2,...,\nu_{max}$. 
 Unlike the harmonic oscillator, the Morse potential has a finite number of bound vibrational 
levels, with $\nu_{max}\simeq2D_e/\omega_e$.

Since the system's dynamics does not change under a constant shift in potential energy then the equation for Morse potential can be rewritten  by adding or subtracting a constant value
\begin{eqnarray}\label{MM1}
V(r)-D_e=D_e (e^{-2a(r-r_e)}-e^{-a(r-r_e)}).
\end{eqnarray}
The Eq.~(\ref{MM1}) approaches zero at infinite $r$ and equals $-D_e$ at its minimum. 
This shows that the Morse potential is the combination of a short-range repulsion and a longer-range attractive tail. The Hamiltonian of a diatomic molecule can be obtained by adding the kinetic term to potential in Eq.~(\ref{MM1}).
In addition, Eq.~(\ref{MM1}) plus kinetic energy can be rewritten in the term of the ladder operators $ \hat S_{\pm} $ and the operator $ \hat S_{0}$~\cite{dong2003controllability}
\begin{eqnarray}\label{Morse Hamiltonian}
\hat H_{p}&=&\hbar\omega_p( \hat S_-\hat S_++\hat S_0),
\end{eqnarray}
where the operators $\hat S_{\pm} $ and $\hat S_{0}$ satisfy the following commutation relations
\begin{eqnarray}\label{commutation}
[\hat S_{+},\hat S_{-}]=2\hat S_{0}, [\hat S_{0},\hat S_{\pm}]=\pm \hat S_{\pm},
\end{eqnarray}
and we have defined the vibrational frequency $\omega_p=\frac{\hbar a^{2}}{2\mu}$ which gets values from GHz~(for $K_2$ molecule) to THz~(for $H_2$ molecule)~\cite{Noggle1996}. 
\begin{figure}[ht]
\centering
\includegraphics[width=3.3in]{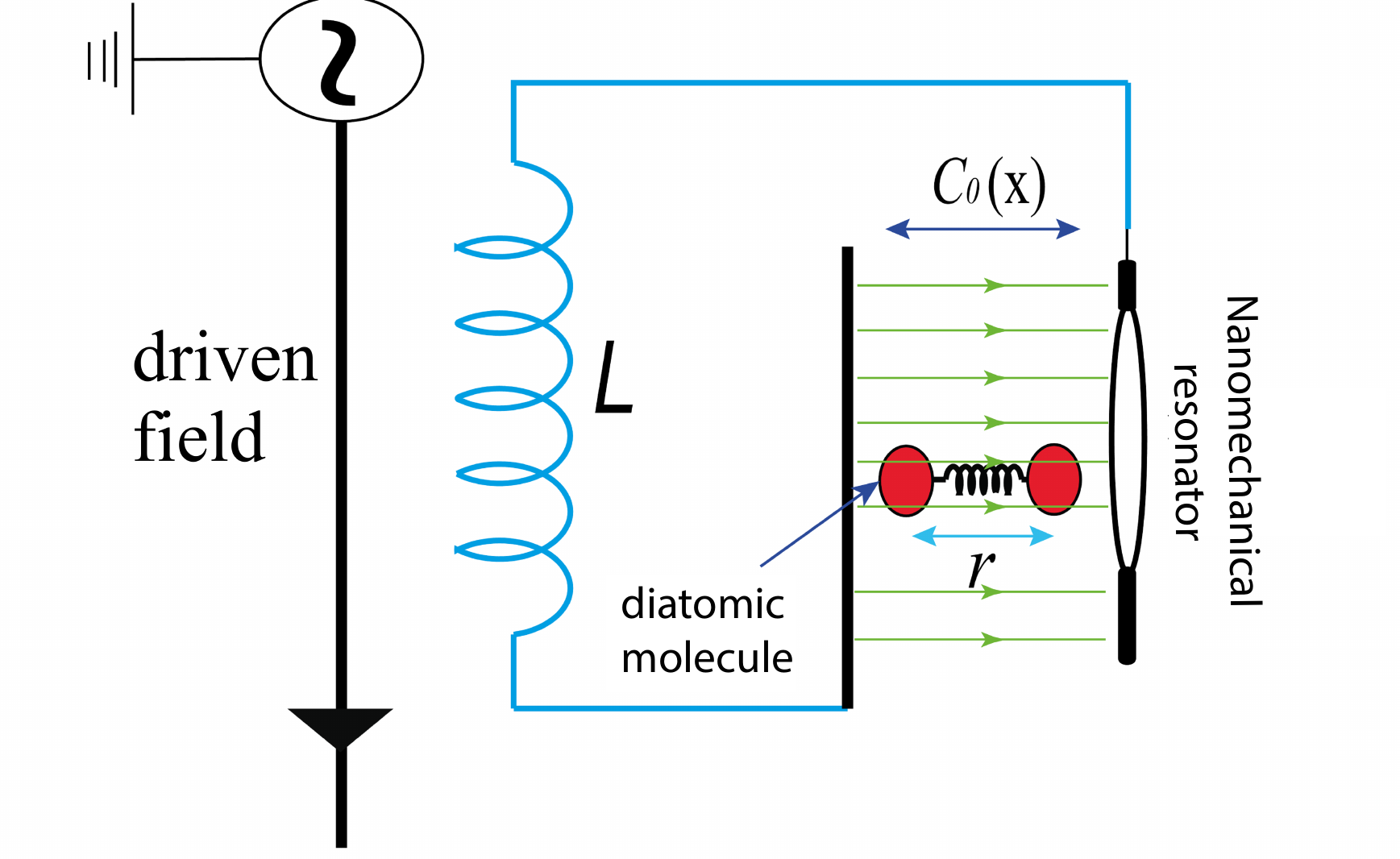}
\caption{(Color online)~The coupling
between a nanomechanical resonator and a single diatomic molecule via microwave cavity field. A single diatomic molecule, described by a Morse potential with transition frequency $\omega_p$, is confined within a microwave cavity in the form of $LC$ circuit. A nanomechanical resonator is also coupled to the central conductor of the microwave cavity via a capacitance $C_0(x)$.  }
\label{fig1}
\end{figure}

\subsection{Hamiltonian of the Hybrid system } 
Now, we are in a position to introduce the hybrid system sketched in Fig.~\ref{fig1}. 
A single diatomic molecule, described by a Morse potential with transition frequency $\omega_p$, is confined within a microwave cavity in the form of $LC$ circuit. A nanomechanical resonator is also coupled to the central conductor of the microwave cavity via a capacitance $C_0(\hat x)=C_0(1-\hat x(t)/d)$. The microwave cavity can be modeled as a single-mode $LC$ 
resonator with frequency $\omega_f=1/\sqrt{LC_0}$, where $C_0$ and $L$ are the overall capacitance and inductance, respectively. 
The nanomechanical resonator is modeled as a harmonic oscillator with frequency $\omega_m$ and effective mass $m$. 
The cavity is driven by an external microwave field at frequency $\omega_0$, where the coherent 
driving of cavity is given by the electric potential $E_d(t)=-i\sqrt{2\hbar\omega_fL}\epsilon_0(e^{i\omega_0t}-e^{-i\omega_0t})$.
Therefore, the Hamiltonian of the hybrid system is described by~\cite{PhysRevA.76.042336}
\begin{equation}\label{Hamiltonian1}
\hat H=\frac{\hat p_x^2}{2m}+\frac{m\omega_m^2 \hat x^2}{2}+\frac{\hat \Phi^2}{2L}+\frac{\hat Q^2}{2C_0(\hat x)}+\hbar\omega_p( \hat S_-\hat S_++\hat S_0)-E_d(t)\hat Q+\hat U_{int}.
\end{equation}
where $\hat x$ and $\hat p_x$ are the canonical position and momentum of the nanomechanical resonator, $\hat \Phi$ and $\hat Q$ denote the canonical coordinates for the microwave cavity with inductance $L$ and capacitance $C_0$, and $\hat U_{int}$ is the interaction potential of the molecule and the microwave cavity mode. 
By expanding the capacitive energy around the equilibrium position of the resonator at $d$ as a Taylor series and by using the annihilation~(creation) operator   $\hat a$ $(\hat a^\dagger)$ of 
the microwave field $([\hat a, \hat a^\dagger] = 1)$, the Hamiltonian~(\ref{Hamiltonian1}) can be rewritten as
\begin{equation}
\hat H=\hbar \omega_f \hat a^\dagger \hat a+\frac{\hbar\omega_m}{2}(\hat{p}^2+\hat{q}^2)+\hbar\omega_p( \hat S_-\hat S_++\hat S_0)+
\hat U_{int}+\frac{1}{2}\hbar G_0 \hat q (\hat a+\hat a^\dagger)^2+i\hbar \epsilon_0(e^{-i\omega_0t}-e^{i\omega_0t})(\hat a+\hat a^\dagger),
\end{equation}
where 
\begin{eqnarray}\label{a}
\hat a&=&\sqrt{\frac{\omega_f L}{2\hbar}}\hat{Q}+\frac{i}{\sqrt{2\hbar\omega_f L}}\hat{\Phi},\\
\hat a^{\dagger}&=&\sqrt{\frac{\omega_f L}{2\hbar}}\hat{Q}-\frac{i}{\sqrt{2\hbar\omega_f L}}\hat{\Phi},\nonumber
\end{eqnarray}
and 
\begin{eqnarray}
G_0&=&\omega_f (\frac{1}{2d}\sqrt{\frac{\hbar}{m\omega_m}}),
\end{eqnarray}
with $\hat q=\sqrt{m \omega_m/\hbar}\hat x$ and $ \hat p=\hat p_x/\sqrt{\hbar m \omega_m}$.

The molecule's dipole can interact with the electric field of the capacitor in the microwave cavity.  The interaction of a radiation field $|\hat{\vec{E}}|=\hat E_x \simeq \hat{Q}/(C_0d)$  with a dipole of the diatomic molecule can be
described by the following Hamiltonian in the dipole approximation~\cite{kielpinski2012quantum,Daniilidis2013}
\begin{equation}\label{U}
\hat U_{int}=e\vec{r}\cdot(E\textbf{x})=\frac{eV_d}{d}(\vec{r}\cdot\textbf{x})\simeq\frac{\hat{Q}}{dC_0} \Big[\sum_{i,j=1,2}\vec{\wp}_{ij} \hat S_{i,j}\Big]\cdot\textbf{x},
\end{equation}
where $V_d$ gives the voltage between the plates of capacitance, $ \textbf{x} $ is the unit vector in the direction of $x$ axis, and $\vec{\wp}_{ij}=e\langle i\vert \vec{r} \vert j\rangle$
 is the electric-dipole transition matrix element of the diatomic molecule. Note that, we approximately ignored the direct interaction of the molecule to the motional degree of freedom of the nanomechanical resonator due to dependence of the capacitance to the position of nanomechanical resonator, i.e., $ C_0(x)\simeq C_0 $. By using Eq.~(\ref{a}), as a result, the molecule-field interaction potential becomes
\begin{eqnarray}
\hat U_{int}=\hbar g(\hat S_++\hat S_-)(\hat a+\hat a^\dagger),
\end{eqnarray}
where $ \hat S_+\equiv\hat S_{2,1} $, $ \hat S_-\equiv\hat S_{1,2} $ and $|g|=\frac{ \hbar(\vec{\wp}_{12}\cdot\textbf{x}) }{d C_0\sqrt{\hbar/2\omega_f L}}$ in which $ |\vec{\wp}_{12}|=|\vec{\wp}_{21}| $.

In an interaction picture with respect to $\hbar\omega_0( \hat a^\dagger \hat a + \hat S_0)$, and after neglecting the terms oscillating at $\pm \omega_0$ and $\pm 2\omega_0$, the Hamiltonian of the system is given by
\begin{equation}\label{Hamiltonian2}
\hat H_I=\hbar\Delta_{0f}\hat a^{\dagger}\hat a+\hbar\Delta_{0p}\hat S_{0}+\hbar\omega_{m}\hat b^{\dagger}\hat b+\hbar\omega_{p}\hat S_{-}\hat S_{+}
-\hbar G_{0}\hat a^{\dagger}\hat a(\hat b+\hat b^{\dagger})+\hbar g(\hat a\hat S_{+}+\hat a^{\dagger}\hat S_{-})+i\hbar\epsilon_{0}(\hat a^{\dagger}-\hat a),
\end{equation}
where $\Delta_{0f}=\omega_{f}-\omega_{0}$ and $\Delta_{0p}=\omega_{p}-\omega_{0}$ are the cavity and Morse potential detuning, respectively. We have used the quantized form of $\hat q$ and $\hat p$ in which $\hat b^\dagger$, $\hat b$($[\hat b,\hat b^\dagger]=1$) 
are the creation and annihilation operators of the nanomechanical resonator excitations, respectively. The first three terms of the Hamiltonian~ (\ref{Hamiltonian2}) 
describe the free evolution energies of the microwave cavity, the diatomic molecule, and the nanomechanical resonator, respectively. 
The term $\hbar\omega_{p}\hat S_{-}\hat S_{+}$ shows the nonlinearity of the diatomic molecule. The fifth term describes the optomechanical coupling between the microwave field and the nanomechanical resonator and the sixth term shows the dipole interaction of the molecule with the microwave cavity mode. The last term, however, 
describes the input driving of the cavity mode by an external microwave field with the coupling strength $\epsilon_0=\sqrt{2\gamma_f P_c/\hbar\omega_f}$, 
where $\gamma_f$ is the decay rate of cavity and $P_c$ is the microwave drive power. It should be noted
that the Hamiltonian, (\ref{Hamiltonian2}), has been written within the
Raman-Nath approximation \cite{schleich2011quantum}, i.e., in the limit when the
atom is allowed only to move over a distance which is
much less than the wavelength of the light. Therefore, in this approximation, one can neglect the kinetic energy and the center of mass motion of the molecule.
 \section{dynamics of the system}
 In this section, first by using master equation we derive the stochastic equations of motion for the hybrid system then by linearizing these equations around their steady state points we find a set of linear equations which describe the dynamics of the tripartite system. 
\subsection{Master equation}
The master equation  for the density operator of the system under Born-Markov and rotating-wave approximations is given by~\cite{gardiner2004quantum}:
\begin{eqnarray}\label{Master}
\frac{d\hat\rho}{dt}=-\frac{i}{\hbar}[\hat H_I,\hat\rho]+L_{f}[\hat\rho]+L_{p}[\hat\rho]+L_{m}[\hat\rho].
\end{eqnarray}
The Liouville terms representing the interaction of the field~($L_{f}$), molecule~($L_{p}$), and mechanical resonator~($L_{m}$) with the heat bath, are given by:
\begin{eqnarray}\label{Liouville}
L_{f}(\hat\rho)&=&(1+\bar{n})\gamma_{f}(2\hat a\hat\rho \hat a^\dagger-\hat\rho \hat a^\dagger \hat a-\hat a^\dagger \hat a \hat\rho)+\bar{n}\gamma_{f}(2\hat a^\dagger\hat\rho \hat a-\hat a\hat a^\dagger\hat\rho-\hat\rho \hat a\hat a^\dagger),\\
L_{p}(\hat\rho)&=&\gamma_{p}(2\hat S_{-}\hat\rho \hat S_{+}-\hat\rho \hat S_{+}\hat S_{-}-\hat S_{+}\hat S_{-} \hat\rho),\\
L_{m}(\hat\rho)&=&(1+n_m)\gamma_{m}(2\hat b\hat\rho \hat b^\dagger-\hat\rho \hat b^\dagger \hat b-\hat b^\dagger \hat b \hat\rho)+n_m\gamma_{m}(2\hat b^\dagger\hat\rho \hat b-\hat b\hat b^\dagger\hat\rho-\hat\rho \hat b\hat b^\dagger).\nonumber
\end{eqnarray}
where  $\gamma_m$ and $\gamma_p$ are the damping rates of the nanomechanical resonator and molecule, respectively. We have also defined $\bar{n}=[\mathrm{exp}(\hbar\omega_f/k_B T)-1]^{-1}$ and $\bar{n}_m=[\mathrm{exp}(\hbar\omega_m/k_B T)-1]^{-1}$, where $ T$ is the temperature of the bath.
 
In order to study dynamics of the system we need to solve the master equation~(\ref{Master}). For this purpose, it is convenient to convert the master equation~(\ref{Master}) to the c-number Fokker-Planck equation
\begin{eqnarray}\label{Fokker}
\frac{\partial P }{\partial t}&=&-\Big\{ \frac{\partial}{\partial\alpha}[-(i\Delta_{0f}+\gamma_f)\alpha - ig\zeta +iG_{0}\alpha(\beta^{*}+\beta) +\epsilon_{0}]+\frac{\partial}{\partial\beta}[-(i\omega_{m}+\gamma_p)\beta +iG_{0}|\alpha|^2]\nonumber\\&+&\frac{\partial}{\partial\zeta_{0}}[-ig(\alpha\zeta^{*}-ig\alpha^{*})\zeta-2\gamma_{p}|\zeta|^{2}]+\frac{\partial}{\partial\zeta}[i\delta_{p}\zeta+2i\omega_{p}\zeta\zeta_{0}+2ig\alpha\zeta_{0}+2\gamma_{p}\zeta_{0}\zeta]+\frac{\partial^{2}}{\partial\alpha^{*}\partial\alpha}[2\gamma_{f}\bar{n}]\\&+&\frac{\partial^{2}}{\partial\beta\partial\beta^{*}}[2\gamma_{m} n_m]+\frac{\partial^{2}}{\partial\alpha\partial\beta}[iG_{0}\alpha]+\frac{\partial^{2}}{\partial\zeta^2}[(i\omega_{p}+\gamma_{p})\zeta^{2}+ig\alpha\zeta]+\frac{\partial^{2}}{\partial\zeta_{0}^2}[-i\frac{g}{2}(\alpha\zeta^{*}-\alpha^{*}\zeta)-\gamma_{p}|\zeta|^2] +c.c.\Big\} P\nonumber
\end{eqnarray}
where $\alpha\equiv \langle \hat a \rangle $, $\beta\equiv \langle \hat b \rangle $, $\zeta\equiv \langle \hat S_- \rangle$, 
$\zeta_0\equiv \langle \hat S_0 \rangle $ and $\delta_p=\omega_p+\omega_0$.

The above Fokker-Planck equation is equivalent to the following set of Ito stochastic differential equations
\begin{eqnarray}\label{Ito}
\dot{\alpha} &=&-(i\Delta_{0f}+\gamma_{f})\alpha -ig\zeta +iG_{0}\alpha(\beta^{*}+\beta) + \epsilon_{0}+\Gamma_{\alpha},\nonumber \\
\dot{\beta}&=&-i\omega_{m}\beta +iG_{0}|\alpha|^2 -\gamma_{m}\beta +\Gamma_{\beta},\\
\dot{\zeta}&=&i\delta_{p}\zeta+2(i\omega_{p}+\gamma_{p})\zeta\zeta_{0}+2ig\alpha\zeta_{0} +\Gamma_{\zeta},\nonumber\\
\dot{\zeta_{0}}&=&-ig\alpha\zeta^*+ig\alpha^{*}\zeta-2\gamma_{p}|\zeta|^2+\Gamma_{\zeta_{0}}\nonumber,
\end{eqnarray}
where $\Gamma_{i}$ are the Gaussian random variables with zero mean and correlations 
\begin{eqnarray}\label{correlation}
\langle\Gamma_{\alpha}(t)\Gamma_{\alpha^{*}}(t^\prime)\rangle&=&4\gamma_{f}\bar{n} \delta(t-t^\prime),\nonumber\\
\langle\Gamma_{\beta}(t)\Gamma_{\beta^{*}}(t^\prime)\rangle&=& 4\gamma_{m}n_m \delta(t-t^\prime),\nonumber\\
\langle\Gamma_{\alpha}(t)\Gamma_{\beta}(t^\prime)\rangle&=&2 iG_{0}\alpha\delta(t-t^\prime),\\
\langle\Gamma_{\zeta}(t)\Gamma_{\zeta}(t^\prime)\rangle&=&2[i\omega_{p}\zeta^{2}+ig\alpha\zeta+\gamma_{p}\zeta^{2}]\delta(t-t^\prime),\nonumber\\
\langle\Gamma_{\zeta_{0}}(t)\Gamma_{\zeta_{0}}(t^\prime)\rangle&=&\left[ -ig(\alpha\zeta^{*}-\alpha^{*}\zeta)-2\gamma_{p}|\zeta|^2\right] \delta(t-t^\prime).\nonumber
\end{eqnarray}

\subsection{Linearization of the Equations of Motion}
The dynamics of the system under study are also determined by the fluctuation-dissipation processes affecting the microwave, molecule and nanomechanical resonator modes. 
They can be taken into account in a fully consistent way by consideration of nonlinear Eqs.~(\ref{Ito}). The system is characterized by  semi-classical steady states
\begin{eqnarray}
\alpha_s &=& \frac{\epsilon_0-ig\zeta_{s}}{i\Delta_f+\gamma_f},\nonumber\\
\zeta_{s}&=&-\frac{ig\alpha_s}{i(\Delta_p/2\zeta_{0s})+\gamma_p },\\
\beta_s &=& \frac{iG_0\vert\alpha_s\vert^2}{i\omega_m-\gamma_m},\nonumber
\end{eqnarray}
where $\alpha_s$, $\zeta_{s}$, and $\beta_{s}$ are the steady states of the cavity, the molecule, and nanomechanical resonator modes, respectively.  $\Delta_f=\Delta_{0f}-G_0(\beta_{s}+\beta_{s}^*)$ shows the effective cavity detuning and $\Delta_p=\delta_p+2\omega_p\zeta_{0s}$ is the effective detuning of the molecule.
 
Eq.~(\ref{Ito}) shows that the dynamics of the system are described by a set of nonlinear equations which can be linearized by writing
 each canonical parameter of the system as a sum of its steady state mean value and a small fluctuation
value~\cite{holmes2009parametric}, i.e., $\alpha=\alpha_{s}+\delta\alpha$, $\beta=\beta_{s}+\delta\beta$, $\zeta=\zeta_{s}+\delta\zeta$, and $\zeta_{0}=\zeta_{0s}+\delta\zeta_{0}$. 
By substituting these operators into Eq.~(\ref{Ito}) and retaining only the first-order terms of fluctuations, the stochastic equations of motion are obtained
\begin{eqnarray}\label{u}
\frac{\partial u}{\partial t} =- \textbf{A}u+\textbf{D}^{1/2}\xi(t),
\end{eqnarray}
where the quadrature fluctuations are defined as $ u=(\delta q,\delta p,\delta X_f,\delta Y_f,\delta x_m,\delta y_m,\delta \zeta_0)^T$ and 
$\delta q\equiv(\delta\beta+\delta\beta^*)/\sqrt{2}$, $\delta p\equiv(\delta\beta-\delta\beta^*)/i\sqrt{2}$, 
$\delta X_f\equiv(\delta \alpha+\delta \alpha^*)/\sqrt{2}$, $\delta Y_f\equiv(\delta\alpha-\delta\alpha^*)/i\sqrt{2}$, 
$\delta x_m\equiv(\delta\zeta^*+\delta\zeta)/\sqrt{2}$, $\delta y_m\equiv(\delta\zeta-\delta\zeta^*)/i\sqrt{2}$. We have also defined the elements of the noise matrix $\xi(t)=(q_{in},p_{in},X_{in},Y_{in},x_{in},y_{in},\Gamma_{\zeta_0})^T$ as 
$q_{in}\equiv(\Gamma_\beta+\Gamma_{\beta^*})/\sqrt{2}$, $p_{in}\equiv(\Gamma_\beta -\Gamma_{\beta^*} )/i\sqrt{2}$, $X_{in}\equiv(\Gamma_\alpha+\Gamma_{\alpha^*})/\sqrt{2}$, $Y_{in}\equiv(\Gamma_\alpha-\Gamma_{\alpha^*})/i\sqrt{2}$, $x_{in}\equiv(\Gamma_{\zeta^*}+\Gamma_{\zeta})/\sqrt{2}$, $y_{in}\equiv(\Gamma_{\zeta}-\Gamma_{\zeta^*})/i\sqrt{2}$.

The drift and diffusion matrices are also given by
\begin{equation}
\textbf{A}=\left(\begin{array}{ccccccc}
    -\gamma_{m} & \omega_{m} &0 &0& 0&0&0 \\
    -\omega_{m} & -\gamma_{m} &2G_0\alpha_s &0 & 0 & 0&0 \\
    0 & 0 & -\gamma_f & \Delta_f &0 & g&0 \\
    2G_0\alpha_s & 0 & -\Delta_f & -\gamma_f &  -g & 0&0\\
    0& 0 &  0 &-G_{0s} & \Gamma_0 & -\Delta_p &K_1\\
    0 &0 & G_{0s}& 0&\Delta_p &\Gamma_0 & K_2\\
    0&0&-g^I&g^R&-\gamma_{p}^{R}&-\gamma_p^I &0
  \end{array}\right),
\label{drift}
\end{equation}
\begin{equation}
\textbf{D}=\left(\begin{array}{ccccccc}
    4\gamma_{m} n_m & 0 &0 &2G_0\alpha_s & 0&0&0 \\
    0 & 4\gamma_{m} n_m &2G_0\alpha_s &0 & 0 & 0&0 \\
    0 & 2G_0\alpha_s & 4\gamma_f \bar n & 0 &0 & 0&0 \\
    2G_0\alpha_s & 0 & 0 & 4\gamma_f\bar n &  0 & 0&0\\
    0& 0 &  0 &0 & P& M &0\\
    0 &0 &0& 0&M &-P & 0\\
    0&0&0&0&0&0 &Q
  \end{array}\right),
\label{diffusion}
\end{equation}
where new variables have defined in Eq.~(\ref{new var}).\\
\section{EFFECTIVE FREQUENCY AND EFFECTIVE DAMPING PARAMETER OF THE NANOMECHANICAL RESONATOR}
This section is devoted to evaluating the effective frequency $\omega_{eff}$ and the effective damping rate $\gamma_{eff}$  of the nanomechanical resonator
 in presence of the diatomic molecule. By solving the linearized  Eq.~(\ref{u}) for the fluctuations in the displacement operator of the nanomechanical resonator, we obtain
\begin{eqnarray}
\delta q(\omega)=\chi(\omega)F_T(\omega),
\end{eqnarray}
where $F_T(\omega)$ is the Fourier transform of the total noises acting on the nanomechanical resonator and $\chi(\omega)$ describes the mechanical effective susceptibility given by
\begin{eqnarray}\label{susceptibility}
\chi(\omega)=\omega_m\Big[ (\gamma_m-i\omega)^{2}+\omega_m^{2}-\frac{(2G_0\alpha_s)^{2}
\omega_m(\Omega_T+i\Gamma_T)(\Omega_{Y^{\prime}}+i\Gamma_{Y^{\prime}})}{(\Omega_X+i\Gamma_X)
(\Omega_Y+i\Gamma_Y)-(\Omega_{X^{\prime}}+i\Gamma_{X^{\prime}})(\Omega_{Y^{\prime}}+i\Gamma_{Y^{\prime}})}\Big]^{-1}.
\end{eqnarray}

The mechanical susceptibility can be considered as susceptibility of an oscillator with an effective resonance frequency and effective damping rate, given by
\begin{eqnarray}\label{effective frequency}
\omega_{eff}=\left[  \gamma_m^{2}+\omega_m^{2}-\frac{(2G_0\alpha_s)^{2}\omega_m \Lambda}{(\Omega_X\Omega_Y-\Gamma_X\Gamma_Y-\Omega_{X^{\prime}}\Omega_{Y^{\prime}}+\Gamma_{X^{\prime}}\Gamma_{Y^{\prime}})^{2}+(\Omega_X\Gamma_Y+\Omega_Y\Gamma_X-\Omega_{X^{\prime}}\Gamma_{Y^{\prime}}-\Omega_{Y^{\prime}}\Gamma_{X^{\prime}})^{2}}\right]  ^{\frac{1}{2}},
\end{eqnarray}
and
\begin{eqnarray}\label{effective loss}
\gamma_{eff}=2\gamma_m +\frac{(2G_0\alpha_s)^{2}\omega_m \Lambda^{\prime}}
{\omega(\Omega_X\Omega_Y-\Gamma_X\Gamma_Y-\Omega_{X^{\prime}}\Omega_{Y^{\prime}}+\Gamma_{X^{\prime}}\Gamma_{Y^{\prime}})^{2}+(\Omega_X\Gamma_Y+\Omega_Y\Gamma_X-\Omega_{X^{\prime}}\Gamma_{Y^{\prime}}-\Omega_{Y^{\prime}}\Gamma_{X^{\prime}})^{2}}
\end{eqnarray}
where the explicit expressions for the parameters
 $\Lambda$, $\Lambda'$, $\Omega_T$, $\Gamma_T$, $\Omega_X$, $\Gamma_X$, $\Omega_{X^{\prime}}$, $\Gamma_{X^{\prime}}$, $\Omega_{Y^{\prime}}$, and $\Gamma_{Y^{\prime}}$ are given in Appendix B. It is evident that in the absence of the molecule, i.e., $\gamma_p=g=0$, and $\Delta_p=\omega_p=0$, the effective frequency Eq.~(\ref{effective frequency})
and the effective damping parameter Eq.~(\ref{effective loss}) reduce to 
the corresponding parameters in the standard optomechanical system~\cite{genes2008ground}. 
\begin{figure}[th]
\centering
\includegraphics[width=3in]{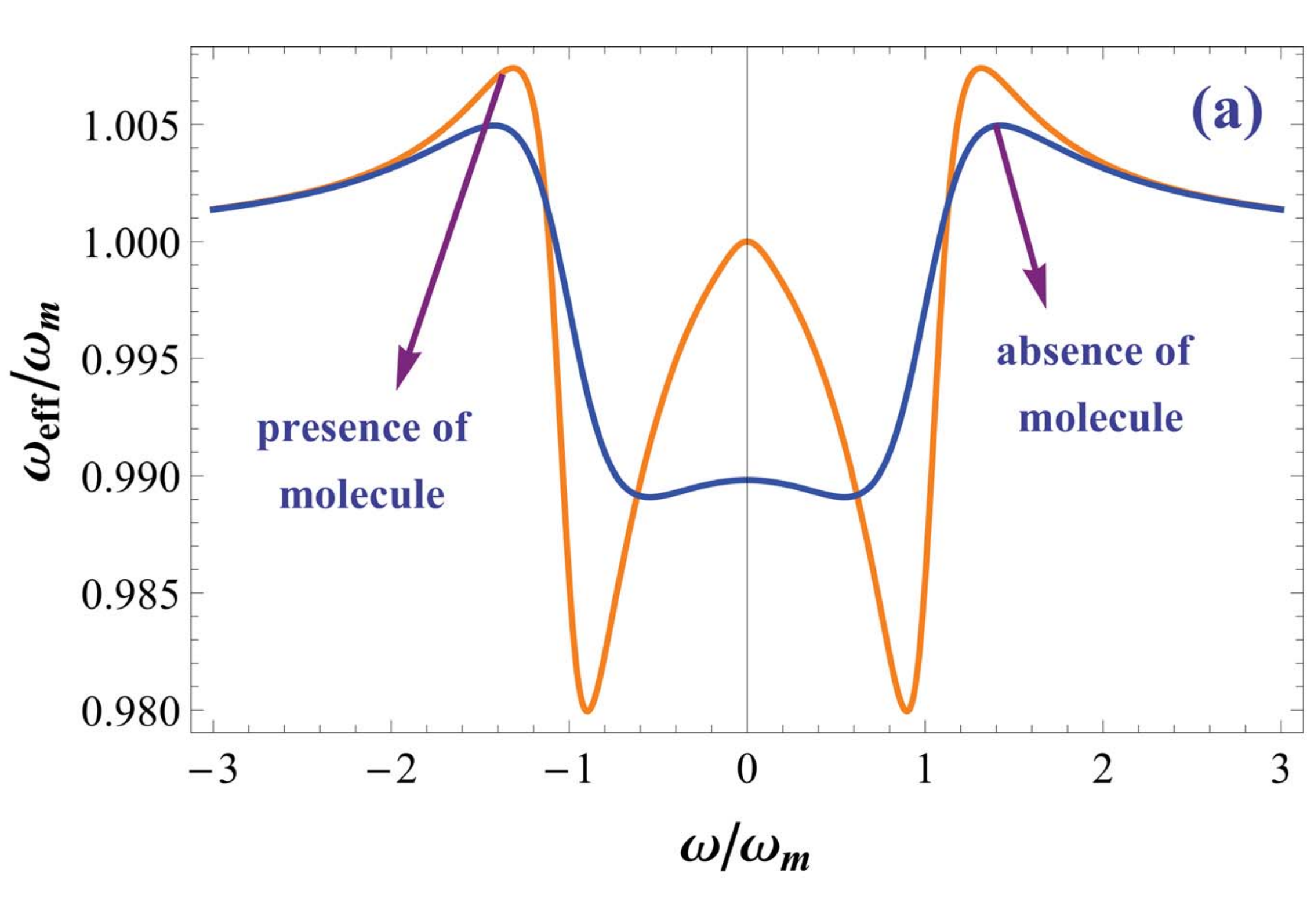}
\includegraphics[width=3in]{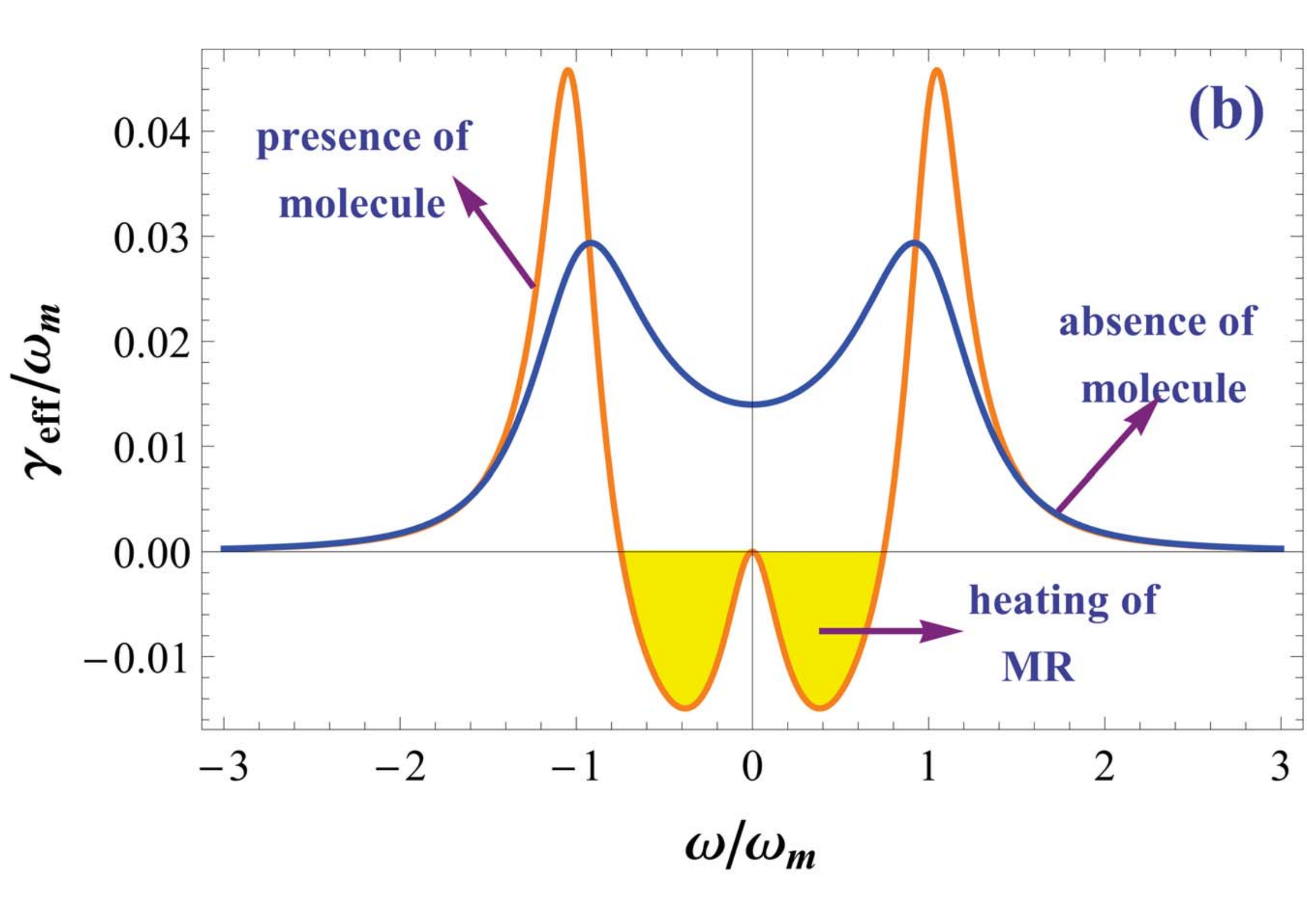}
\caption{(Color online)
(a)~The normalized effective frequency $\omega_{eff}/\omega_m$
as a function of the normalized system response 
frequency $\omega/\omega_m$ at $\Delta_p/|\zeta_{0s}|=-\omega_m$ in presence and absence of the molecule. (b)~The normalized effective damping parameter $\gamma_{eff}/\omega_m$
as a function of the normalized system response 
frequency $\omega/\omega_m$ at $\Delta_p/|\zeta_{0s}|=-\omega_m$ in presence and absence of the molecule. The cavity damping is assumed to be $\gamma_f=0.4 \omega_m$, while the other parameters are $P_c=5$mW, $\omega_m/2\pi=10$MHz, $m=10$ng, $Q=50\times 10^4$, $d=100$nm, $\beta=0.013$, $\gamma_a=0.8\omega_m$ and $g=5$kHz.}
\label{fig3}
\end{figure}
\begin{figure}[th]
\centering
\includegraphics[width=3in]{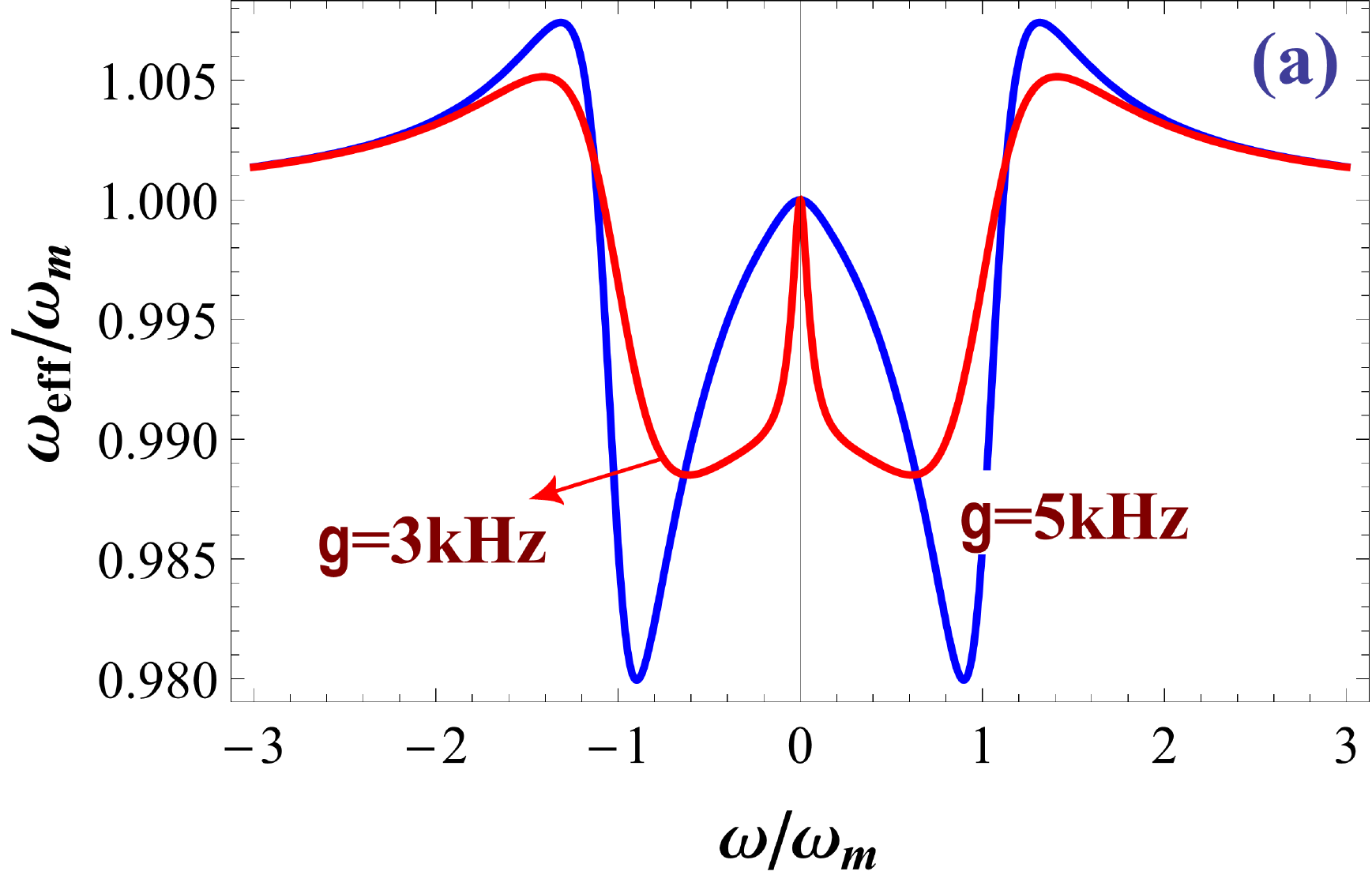}
\includegraphics[width=3in]{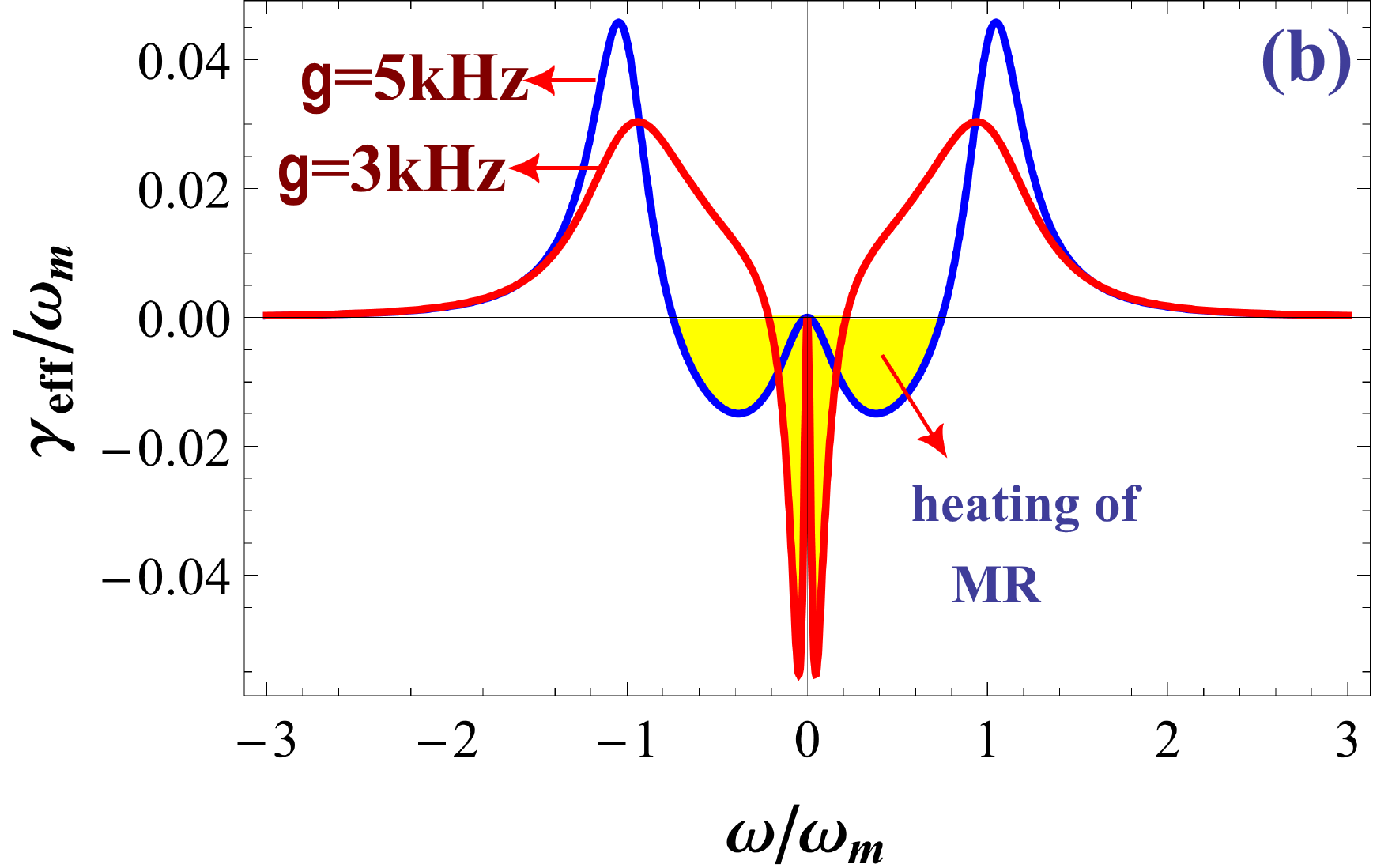}
\caption{(Color online)
(a)~The normalized effective frequency $\omega_{eff}/\omega_m$ 
as a function of the normalized system response 
frequency $\omega/\omega_m$ for two different values of molecule-field coupling constant $ g=3 $kHz and $ g=5 $kHz. (b)~The normalized effective damping parameter $\gamma_{eff}/\omega_m$
as a function of the normalized system response 
frequency $\omega/\omega_m$ for two different values of molecule-field coupling constant $ g=3 $kHz and $ g=5 $kHz. Here, we have assumed $\Delta_p/|\zeta_{0s}|=-\omega_m$ and the other parameters are the same as Fig.~\ref{fig3}.}
\label{fig44}
\end{figure}

Fig.~\ref{fig3} shows the normalized effective frequency 
as a function of the normalized system response 
frequency $\omega/\omega_m$ at $\Delta_p/|\zeta_{0s}|=-\omega_m$ in presence and absence of the molecule in a case of $ \zeta_0=-1 $. The cavity damping is assumed to be $\gamma_f=0.4 \omega_m$, while the other parameters are~\cite{teufel}:  $P_c=5$mW, $\omega_m/2\pi=10$MHz, $m=10$ng, $Q=50\times 10^4$, $d=100$nm, $\beta=0.013$, $\gamma_a=0.8\omega_m$ and $g=5$kHz~\cite{rabl2006hybrid}. We see that presence of the molecule potentially increases the effective frequency and damping parameter of the nanomechanical resonator. Further information can be found in Fig.~\ref{fig44}, which shows the influence of the molecule-field coupling rate $ g $ on the effective frequency and damping parameter. This figure confirms that, by increasing the parameter $ g $, both parameters $ \omega_{eff} $ and $\gamma_{eff}$ increase. 
 \begin{figure}[th]
\centering
\includegraphics[width=3in]{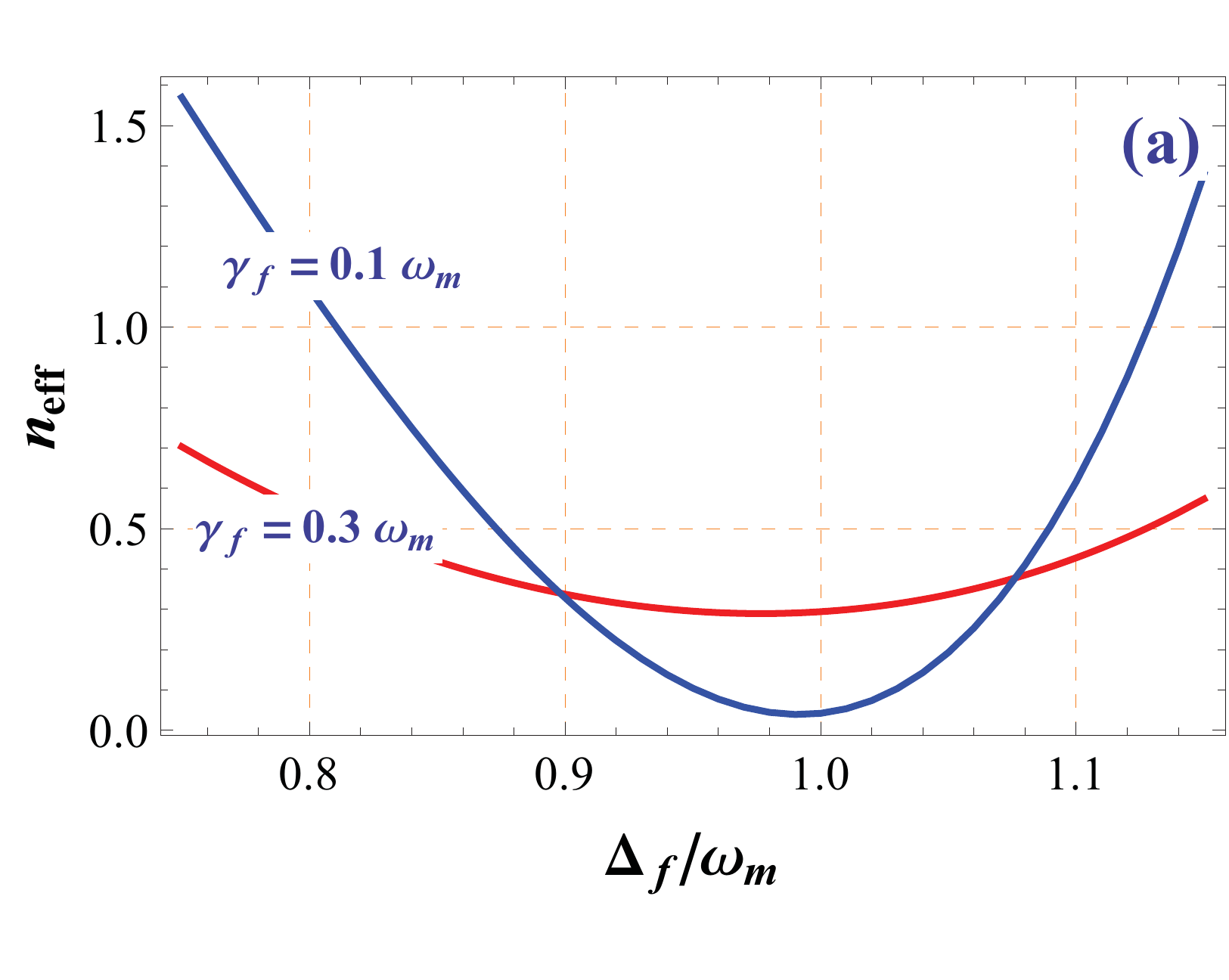}
\includegraphics[width=3in]{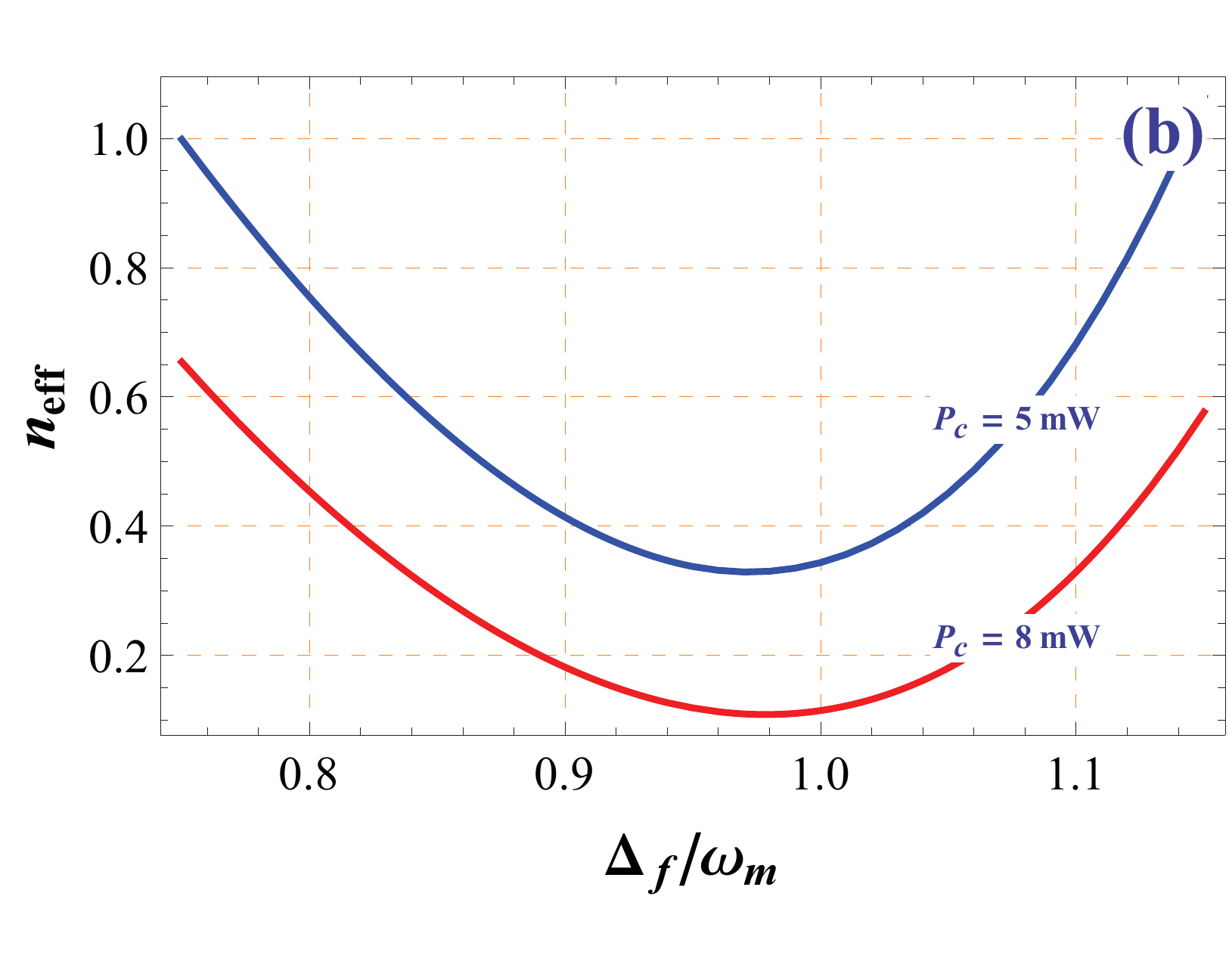}
\caption{(Color online)
(a)~The effective mean excitation $n_{eff}$ versus the normalized effective detuning $ \Delta_f/\omega_m $ for different values of the cavity damping rate $\gamma_f=0.1 \omega_m$ and $\gamma_f=0.3 \omega_m$. (b)~The effective mean excitation $n_{eff}$ versus the normalized effective detuning $ \Delta_f/\omega_m $ for different values of the  driven power $P_c=5$mW and $P_c=8$mW. The molecular detuning has been fixed at $\Delta_p/|\zeta_0|=-\omega_m$, while the other parameters are  $\omega_m/2\pi=10$MHz, $m=10$ng, $Q=50\times 10^4$, $d=100$nm, $\beta=0.013$, $g=10$kHz and the reservoir temperature is $T = 0.2 K$.}
\label{fig4}
\end{figure}
\section{GROUND STATE COOLING OF THE NANOMECHANICAL RESONATOR}
In this section, we study the ground state cooling of the nanomechanical resonator coupled to the diatomic molecule in the steady state. We note that the current system is stable and reaches a steady
state after a transient time if all the eigenvalues of the
drift matrix $\textbf{A}$ have negative real part. These stability
conditions can be obtained by using the Routh-Hurwitz criteria~\cite{gradshteyn1980table}. 
The mean energy of the nanomechanical resonator in the steady state is
\begin{equation}
U=\frac{\hbar\omega_m}{2}\Big[\langle\delta q^{2}\rangle+\langle\delta p^{2}\rangle\Big]=\frac{\hbar\omega_m}{2}(V_{11}+V_{22})\equiv \hbar \omega_m(n_{eff}+\frac{1}{2})\nonumber,
\end{equation}
where $V_{11}$ and $V_{22}$ are the first and second diagonal components of the stationary correlation matrix 
\begin{eqnarray}
V_{ij}=\int_{0}^{\infty}ds\int_{0}^{\infty}ds^\prime M_{ik}(s)M_{jl}(s^\prime)D_{kl}(s-s^\prime),
\end{eqnarray}
where $\textbf{M}(t)=\exp(\textbf{A}t)$ and $\textbf{D}(s-s^\prime)$ is the diffusion matrix given by Eq.~(\ref{diffusion}). It is worth mentioning that, the nanomechanical resonator reaches its ground state if $n_{eff}\simeq 0$ or $V_{11}\equiv\langle\delta q^{2}\rangle\simeq1/2$ 
and $V_{22}\equiv\langle\delta p^{2}\rangle\simeq1/2$. 

In Fig.~\ref{fig4}(a) we have plotted the effective number of vibrational excitations $n_{eff}$ as a function of the normalized cavity detuning $\Delta_f/\omega_m$ 
for two different values of the cavity damping rate
$\gamma_f$ in $\Delta_p/|\zeta_0|=-\omega_m$. The microwave cavity is assumed to work in frequency $\omega_f/2\pi=10$GHz which is driven by a microwave source with power $P_c=8$mW. We also have considered molecule $K_2$  with $\omega_e=17 $THz, $D_e=8.1\times10^{-20} $J, $\omega_p/2\pi=90 $GHz~\cite{Noggle1996}, and $g=10$kHz. Fig.~\ref{fig4} (a) shows that decreasing the cavity damping improves the cooling process for nanomechanical resonator.  However, Fig.~\ref{fig4}(b) describes the effect of the laser power $P_c$ on the ground state cooling of nanomechanical resonator. It is evident that  
the lower nanomechanical resonator excitation is obtained in the higher values of the input power. 
\begin{figure}[th]
\centering
\includegraphics[width=2.8in]{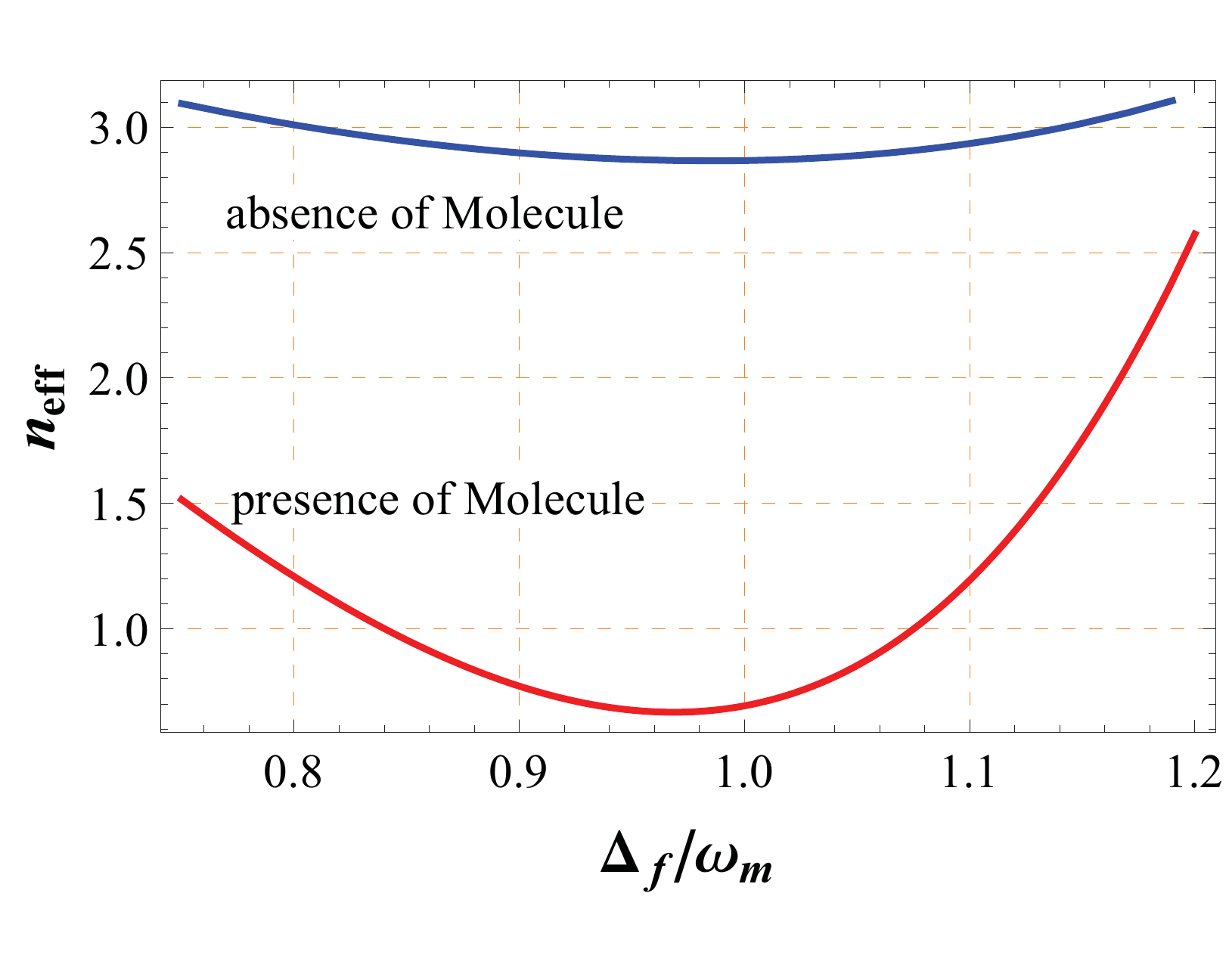}
\caption{(Color online)
Comparing the effective mean excitation $n_{eff}$ in presence and absence of the molecule for $ g=10 $kHz. The other parameters
are the same as Fig.~\ref{fig4}.}
\label{fig5}
\end{figure}
\begin{figure}[th]
\centering
\includegraphics[width=2.8in]{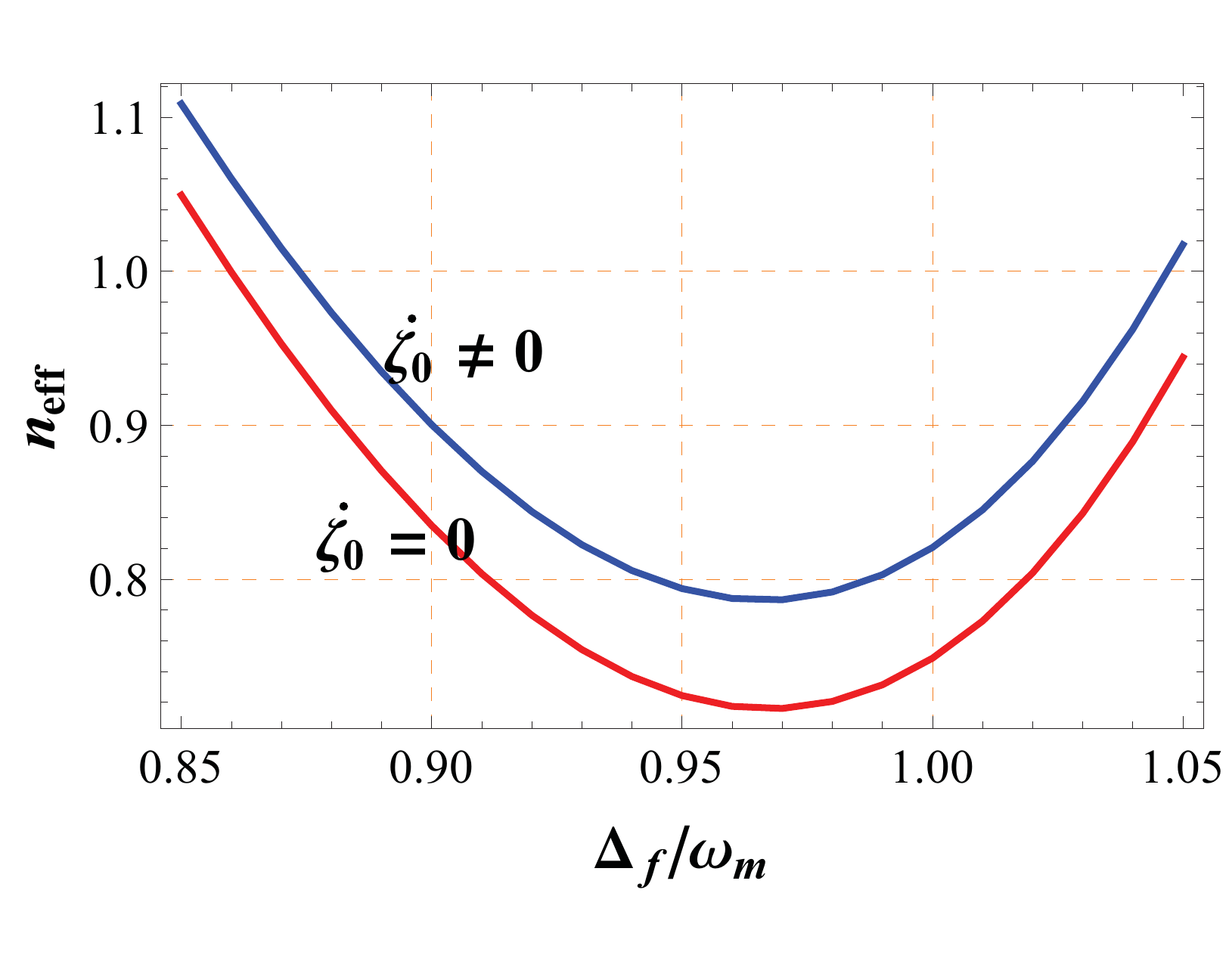}
\caption{(Color online)
The effective mean excitation $n_{eff}$ for two different regimes $\dot{\zeta}_0=0$ and $\dot{\zeta}_0\neq0$ with $ g=10 $kHz. The other parameters
are the same as Fig.~\ref{fig4}.}
\label{fig6}
\end{figure}
\begin{figure}[th]
\centering
\includegraphics[width=3.1in]{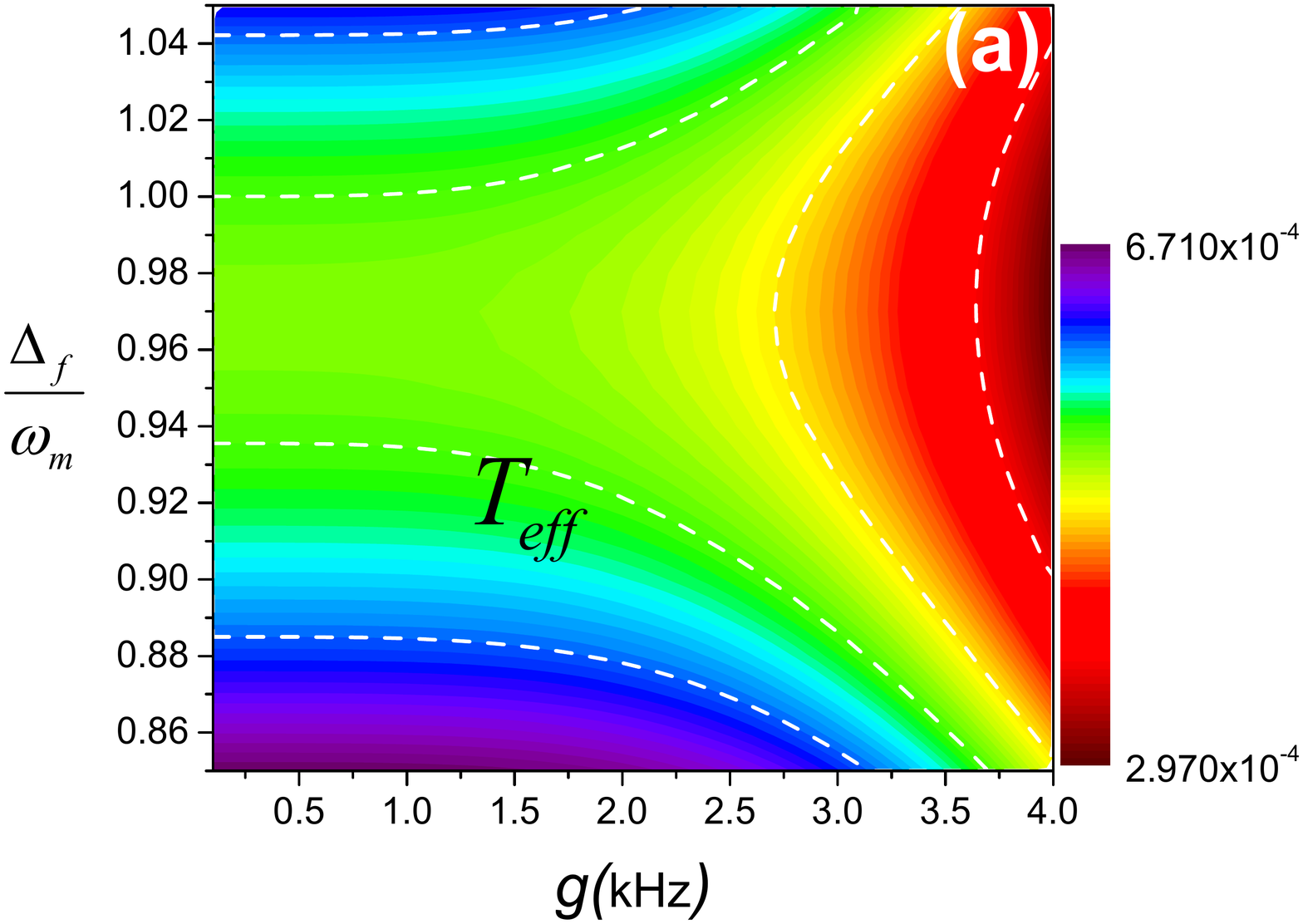}
\includegraphics[width=3.1in]{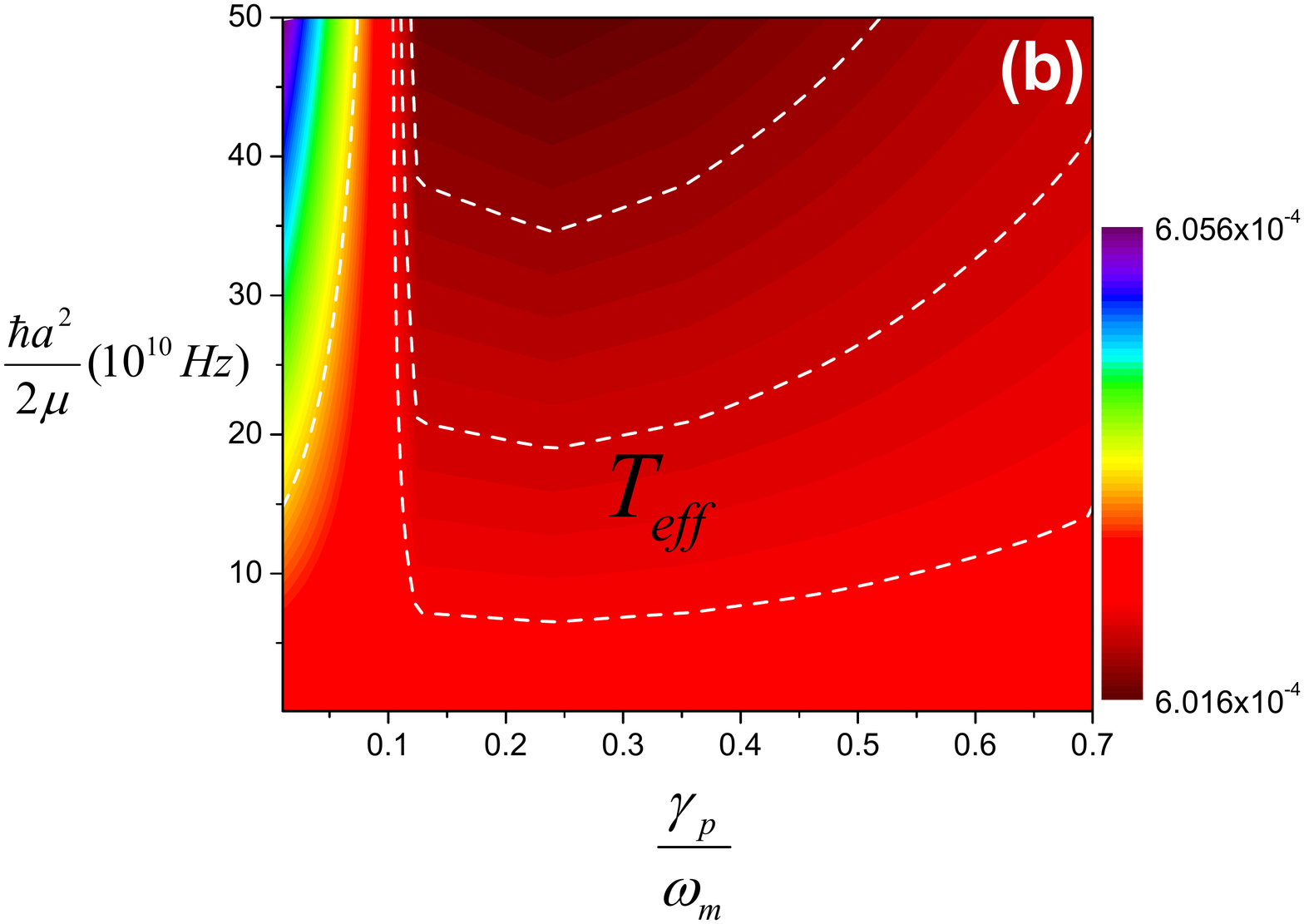}
\caption{(Color online)
 (a)~The effective temperature $T_{eff}$ versus normalized cavity field detuning $ \Delta_f/\omega_m $ and molecule-field coupling contestant $ g $. (b)~The effective temperature $T_{eff}$ versus the molecule nonlinearity parameter $ \hbar a^2/2\mu $ and the normalized molecule damping rate $ \gamma_p/\omega_m $.  The other parameters
are the same as Fig.~\ref{fig4}.}
\label{fig7}
\end{figure}
A more interesting situation is depicted in Fig.~\ref{fig5} which compares $n_{eff}$ in presence and absence of the molecule. This figure reveals that presence of the molecule potentially could improve the cooling process for the nanomechanical resonator. This figure is depicted for situation $ \dot{\zeta}_0 \neq 0$ which corresponds to the last line of Eq.~(\ref{Ito}). However, the excitation probability of a single molecule is usually assumed to be low i.e., $\zeta_0= \mathrm{cte}$  or $\dot{\zeta}_0\simeq0$. This is valid in the low molecular excitation limit, i.e., when
the molecule is initially prepared in its ground state and interaction with the cavity field does not effectively change the molecule excitation. This
means that the single-molecule excitation probability should be
much less than 1, i.e., $ g^2\ll \Delta_p^2+\gamma_p^2 $. In this regime, one can assume $\dot{\zeta}_0\simeq0$, which is equivalent to ignoring the last differential equation in Eq.~(\ref{Ito}).
 In Fig.~\ref{fig6} we have plotted $ n_{eff} $ versus the normalized detuning $ \Delta_f/\omega_m $ for two different scenarios $\dot{\zeta}_0\simeq0$ and $\dot{\zeta}_0\neq0$. It is interesting that, in the case of the low molecular excitation limit i.e., $\dot{\zeta}_0\simeq0$, one can reach lower temperature for nanomechanical resonator. In principle, this approximation is identical with the bosonization method proposed for atoms~\cite{Genes2008}. 

Now, we investigate the effect of the molecule parameters on the cooling of nanomechanical resonator in the low molecular excitation regime. The effective temperature of the nanomechanical resonator $ T_{eff}=\frac{\hbar\omega_m}{\mathrm{k_B} \mathrm{ln}(1+1/n_{eff})} $ versus the normalized cavity detuning $ \Delta_f/\omega_m $ and the molecule-field coupling rate $ g $ is depicted in Fig.~\ref{fig7}(a). As expected, the stronger coupling between the intracavity mode and the molecule leads to a lower temperature for nanomechanical resonator.  Figure~\ref{fig7}(b) shows $ T_{eff} $ versus the molecule damping and parameter $  \omega_p=\hbar a^2/2\mu $ in the fixed cavity detuning $ \Delta_f=\omega_m $. This is interesting to mentioned that for $ \gamma_p\gtrsim 0.12 \omega_m $ by increasing the parameter $ \omega_p $, which indicates the strength of molecule nonlinearity,  the effective temperature decreases.
However, the lowest nanomechanical resonator temperature is approached around $ \gamma_p\simeq 0.25 \omega_m$. On the other hand, for $ \gamma_p\lesssim 0.12 \omega_m $, increasing the parameter  $\omega_p$ suppresses the cooling process of the nanomechanical resonator.

\begin{figure}[th]
\centering
\includegraphics[width=3.2in]{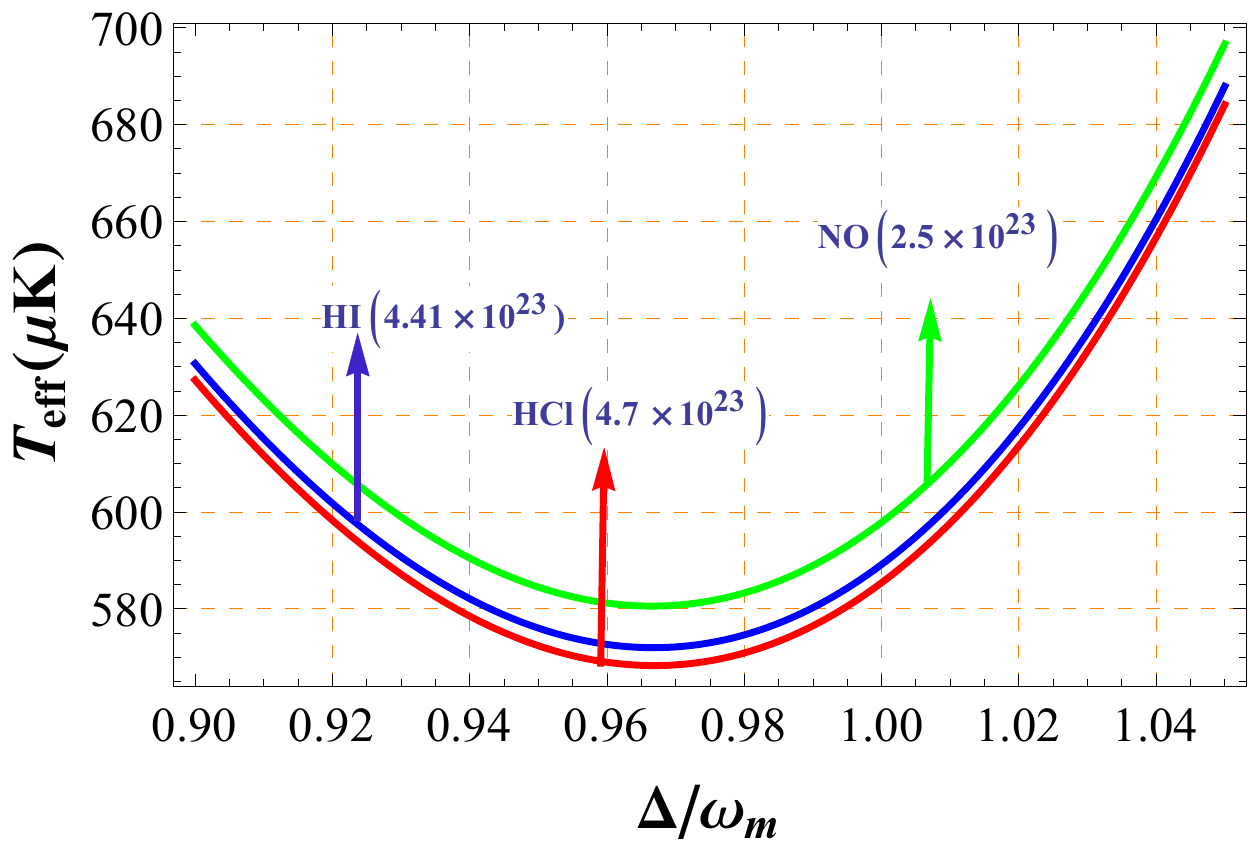}
\caption{(Color online) The effective temperature  $T_{eff}$ versus the normalized effective detuning  $ \Delta/\omega_m $ for three different molecules $\mathrm{HCl}$, $\mathrm{HI}$, and $\mathrm{NO}$. Here,
 $\omega_e/\sqrt{2D_e}=4.7 \times 10^{23}$ for $\mathrm{HCl}$, $\omega_e/\sqrt{2D_e}=4.41 \times 10^{23}$ for $\mathrm{HI}$, and $\omega_e/\sqrt{2D_e}=2.5 \times 10^{23}$ for $\mathrm{NO}$. The other parameters are the same as Fig.~\ref{fig4}.
}
\label{fig8}
\end{figure}

 Finally, Fig.~\ref{fig8} shows the effective temperature $ T_{eff} $ versus $ \Delta_f/\omega_m $ in the presence of three different types of the diatomic molecules $\mathrm{HCl}$, $\mathrm{HI}$, and $\mathrm{NO}$ where each molecule is indicated by parameter $\omega_e/\sqrt{2D_e}$. Here, $\omega_e/\sqrt{2D_e}=4.7 \times 10^{23}$ for $\mathrm{HCl}$, $\omega_e/\sqrt{2D_e}=4.41 \times 10^{23}$ for $\mathrm{HI}$, and $\omega_e/\sqrt{2D_e}=2.5 \times 10^{23}$ for $\mathrm{NO}$~\cite{Noggle1996}.  It is evident that by increasing the value of $\omega_e/\sqrt{2D_e}$, which increases the width of potential well for constant $\mu$, we achieve a lower temperature for the nanomechanical resonator. 

We note that one of the realizations of the diatomic molecules are in DNA molecules~\cite{zdravkovic2012morse}. It is well-known that a DNA molecule consists of two compatible chains. Chemical bonds between neighbouring nucleotides belonging to the same strands are strong covalent bonds, while nucleotides at a certain site $  n$, belonging to
the different strands, are connected through weak hydrogen interaction~\cite{peyrard1989statistical,dauxois1991dynamics}.  Interaction between nucleotides at the same site belonging to different strands is modelled by a mechanical resonator energy.

\section{CONCLUSIONS}
In this paper, we have proposed a theoretical scheme for
realization of tripartite coupling among a
single mode of the microwave cavity, a single diatomic molecule, and the vibrational mode of
a nanomechanical resonator. We have shown that, by describing the diatomic molecule with a Morse potential, a type
of tripartite molecule-nanomechanical resonator-field coupling can be manifested.
 The dynamics of the hybrid system is studied by using the Fokker-Planck equation. We
have focused our attention on the steady state of the system
and, in particular, on the stationary quantum fluctuations of the
system. By solving the linearized dynamics around the classical
steady state we have found the effective frequency and the effective
damping parameter of the nanomechanical resonator. We have seen that, in an experimentally accessible
parameter regime, presence of the molecule modifies the effective frequency and the damping rate of the nanomechanical resonator. Moreover, we have studied the cooling of the nanomechanical resonator and have shown that presence of the diatomic molecule improves the efficiency of the cooling mechanism for the nanomechanical resonator.  We have found that by increasing the coupling constant between the molecule and cavity field, the effective temperature of the nanomechanical resonator decreases. We have also discussed about the effect of molecular parameters on the temperature of the nanomechanical resonator. We have shown that, by increasing the molecule parameter $\omega_e/\sqrt{2D_e}$, one can reach lower temperatures for the nanomechanical resonator. The realization of such a scheme will
open new opportunities for coupling between nanomechanical resonators and diatomic-like molecules such as DNA.
\section*{Acknowledgements} 
M.E.A, H.Y, and M.A.S wish to thank the Office of Graduate
Studies of The University of Isfahan for their support. The work of S.B has been supported by the Alexander von Humboldt foundation. 

\appendix
\section{Definition of  variables in Eqs.~(\ref{drift}) and (\ref{diffusion})}
\begin{eqnarray}\label{new var}
G_{0s}&=&2g\zeta_{0s},\,\, \Gamma_0=2\gamma_p\zeta_{0s},\, g^R=\sqrt{2}g \mathrm{Re}\zeta_{s},\nonumber\\
g^I&=&\sqrt{2}g\mathrm{Im}\zeta_{s},\,\, \gamma_p^R =2\sqrt{2}\gamma_p \mathrm{Re}\zeta_{s},\nonumber\\
\gamma_p^I &=&2\sqrt{2}\gamma_p \mathrm{Im}\zeta_{s}+\sqrt{2}g\alpha_s,\\
K_1&=&\sqrt{2}\left[ 2\gamma_p \mathrm{Re} \zeta_{s}-\mathrm{Im}(2\omega_p\zeta_{s}+2g\alpha_s)\right] ,\nonumber\\
K_2&=&\sqrt{2}\left[ 2\gamma_p \mathrm{Im} \zeta_{s}+ \mathrm{Re}(2\omega_p\zeta_{s}+2g\alpha_s)\right] ,\nonumber\\
P&=&\langle\Gamma_{\zeta}\Gamma_{\zeta}\rangle +\langle\Gamma_{\zeta^{*}}\Gamma_{\zeta^{*}}\rangle,\nonumber\\
M&=&-i\left[ \langle\Gamma_{\zeta}\Gamma_{\zeta}\rangle -\langle\Gamma_{\zeta^{*}}\Gamma_{\zeta^{*}}\rangle\right],
\,\,Q=\langle\Gamma_{\zeta_{0}}\Gamma_{\zeta_{0}}\rangle.\nonumber
\end{eqnarray}
\section{Definition of  variables in Eqs.~(\ref{effective frequency}) and (\ref{effective loss})}
\begin{eqnarray}
\Lambda &=&\left( \Omega_{T}\Omega_{Y^\prime} - \Gamma_{T}\Gamma_{Y^\prime}\right) (\Omega_{X}\Omega_{Y} - \Gamma_{X}\Gamma_{Y} -\Omega_{X^\prime}\Omega_{Y^\prime}+\Gamma_{X^{\prime}}\Gamma_{Y^{\prime}})\nonumber\\&+&(\Omega_{T}\Gamma_{Y^\prime}+\Gamma_{T}\Omega_{Y^\prime}+\Gamma_{T}\Omega_{Y^\prime})(\Omega_{X}\Gamma_{Y} +\Omega_{Y}\Gamma_{X} -\Omega_{X^\prime}\Gamma_{Y^\prime}-\Omega_{Y^\prime}\Gamma_{X^\prime}),\nonumber\\
\Lambda^{\prime}&=&(\Omega_{T}\Gamma_{Y^\prime}+\Omega_{Y}\Gamma_{T})(\Omega_{X}\Omega_{Y} -\Gamma_{X}\Gamma_{Y} -\Omega_{X^\prime}\Omega_{Y^\prime}+\Gamma_{X^\prime}\Gamma_{Y^\prime})\nonumber\\&-&(\Omega_{T}\Omega_{Y^\prime} -\Gamma_{T}\Gamma_{Y^\prime})(\Omega_{X}\Omega_{Y} +\Gamma_{X}\Gamma_{Y})-\Omega_{X^\prime}\Omega_{Y^\prime}-\Omega_{Y^\prime}\Gamma_{X^\prime}),\nonumber\\
\Omega_{T} &=&\Omega_p^2+\Gamma_0^2 -\left( \frac{K_2\gamma_p^I}{\omega}-\omega\right) \left( \frac{K_1\gamma_p^R}{\omega}-\omega\right)+\left( \frac{K_2 \gamma_p^R}{\omega}\right) \left( \frac{K_1\gamma_p^I}{\omega}\right),\nonumber\\
\Gamma_{T} &=& \frac{K_1\gamma_p^I \Omega_p}{\omega}-\frac{K_2\gamma_p^R\Omega_p}{\omega}-\Gamma_0\left( \frac{K_1\gamma_p^R}{\omega}-\omega\right)-\Gamma_0\left(\frac{K_2\gamma_p^I}{\omega}-\omega\right),\nonumber\\
\Omega_X &=&\gamma_f\Omega_T +\omega\Gamma_T- gG_{0s}\Gamma_0 + g \left( \frac{K_2\gamma_p^R}{\omega}\right) \left( \frac{K_1g^I}{\omega}\right)+g\left(\frac{K_2 g^I}{\omega} \right)\left( \frac{K_1\gamma_p^R}{\omega}-\omega\right),\\
\Gamma_X &=&\gamma_f\Gamma_T -\omega\Omega_T +g\Omega_p\left( \frac{K_1g^I}{\omega}\right)-gG_{0s}\left( \frac{K_1\gamma_p^R}{\omega}-\omega\right)+g\Gamma_0\left( \frac{K_2g^I}{\omega}\right),\nonumber\\
\Omega_Y &=&\gamma_f\Omega_T + \omega\Gamma_T- g G_{0s}\Gamma_0+ g\left( \frac{K_1\gamma_p^I}{\omega}\right) \left( \frac{K_2g^R}{\omega}\right)+g\left(\frac{K_1 g^R}{\omega} \right)\left( \frac{K_2\gamma_p^I}{\omega}-\omega\right),\nonumber\\
\Gamma_Y &=&\gamma_f\Gamma_T -\omega\Omega_T +g\Omega_p\left( \frac{K_2g^R}{\omega}\right)-gG_{0s}\left( \frac{K_2\gamma_p^I}{\omega}-\omega\right)+g\Gamma_0\left( \frac{K_1g^R}{\omega}\right),\nonumber\\
\Omega_{X^\prime}&=&-\Delta\Omega_T+ g \Omega_p G_{0s}+g\left( \frac{K_1\gamma_p^I}{\omega}\right) \left( \frac{K_2g^I}{\omega}\right)- g\left( \frac{K_1g^I}{\omega}\right) \left( \frac{K_2\gamma_p^I}{\omega}-\omega\right) ,\nonumber\\
\Gamma_{X^\prime}&=&-\Delta\Gamma_T- g\Omega_p\left( \frac{K_2 g^I}{\omega}\right)+gG_{0s}\left(\frac{K_1\gamma_p^I}{\omega}\right)-g\Gamma_0\left( \frac{K_1g^I}{\omega}\right),\nonumber\\
 \Omega_{Y^\prime}&=&\Delta\Omega_T- g \Omega_p G_{0s}+g\left( \frac{K_2g^R}{\omega}\right) \left( \frac{K_1\gamma_p^R}{\omega}-\omega\right)- g\left( \frac{K_1g^R}{\omega}\right) \left( \frac{K_2\gamma_p^R}{\omega}\right) ,\nonumber\\
\Gamma_{Y^\prime}&=&\Delta\Gamma_T+g\Omega_p\left( \frac{K_1 g^R}{\omega}\right)+gG_{0s}\left(\frac{K_2\gamma_p^R}{\omega}\right)
-g\Gamma_0\left( \frac{K_2g^R}{\omega}\right).\nonumber
\end{eqnarray}


\bibliography{sci}

\begin{thebibliography}{44}%
\makeatletter
\providecommand \@ifxundefined [1]{%
 \@ifx{#1\undefined}
}%
\providecommand \@ifnum [1]{%
 \ifnum #1\expandafter \@firstoftwo
 \else \expandafter \@secondoftwo
 \fi
}%
\providecommand \@ifx [1]{%
 \ifx #1\expandafter \@firstoftwo
 \else \expandafter \@secondoftwo
 \fi
}%
\providecommand \natexlab [1]{#1}%
\providecommand \enquote  [1]{``#1''}%
\providecommand \bibnamefont  [1]{#1}%
\providecommand \bibfnamefont [1]{#1}%
\providecommand \citenamefont [1]{#1}%
\providecommand \href@noop [0]{\@secondoftwo}%
\providecommand \href [0]{\begingroup \@sanitize@url \@href}%
\providecommand \@href[1]{\@@startlink{#1}\@@href}%
\providecommand \@@href[1]{\endgroup#1\@@endlink}%
\providecommand \@sanitize@url [0]{\catcode `\\12\catcode `\$12\catcode
  `\&12\catcode `\#12\catcode `\^12\catcode `\_12\catcode `\%12\relax}%
\providecommand \@@startlink[1]{}%
\providecommand \@@endlink[0]{}%
\providecommand \url  [0]{\begingroup\@sanitize@url \@url }%
\providecommand \@url [1]{\endgroup\@href {#1}{\urlprefix }}%
\providecommand \urlprefix  [0]{URL }%
\providecommand \Eprint [0]{\href }%
\providecommand \doibase [0]{http://dx.doi.org/}%
\providecommand \selectlanguage [0]{\@gobble}%
\providecommand \bibinfo  [0]{\@secondoftwo}%
\providecommand \bibfield  [0]{\@secondoftwo}%
\providecommand \translation [1]{[#1]}%
\providecommand \BibitemOpen [0]{}%
\providecommand \bibitemStop [0]{}%
\providecommand \bibitemNoStop [0]{.\EOS\space}%
\providecommand \EOS [0]{\spacefactor3000\relax}%
\providecommand \BibitemShut  [1]{\csname bibitem#1\endcsname}%
\let\auto@bib@innerbib\@empty
\bibitem [{\citenamefont {Aspelmeyer}\ \emph {et~al.}(2014)\citenamefont
  {Aspelmeyer}, \citenamefont {Kippenberg},\ and\ \citenamefont
  {Marquardt}}]{Aspelmeyer}%
  \BibitemOpen
  \bibfield  {author} {\bibinfo {author} {\bibfnamefont {M.}~\bibnamefont
  {Aspelmeyer}}, \bibinfo {author} {\bibfnamefont {T.~J.}\ \bibnamefont
  {Kippenberg}}, \ and\ \bibinfo {author} {\bibfnamefont {F.}~\bibnamefont
  {Marquardt}},\ }\href@noop {} {\bibfield  {journal} {\bibinfo  {journal}
  {Rev. Mod. Phys.}\ }\textbf {\bibinfo {volume} {86}},\ \bibinfo {pages}
  {1391} (\bibinfo {year} {2014})}\BibitemShut {NoStop}%
\bibitem [{\citenamefont {Chen}(2013)}]{Chen}%
  \BibitemOpen
  \bibfield  {author} {\bibinfo {author} {\bibfnamefont {Y.}~\bibnamefont
  {Chen}},\ }\href@noop {} {\bibfield  {journal} {\bibinfo  {journal} {J. Phys.
  B: At. Mol. Opt. Phys.}\ }\textbf {\bibinfo {volume} {46}},\ \bibinfo {pages}
  {104001} (\bibinfo {year} {2013})}\BibitemShut {NoStop}%
\bibitem [{\citenamefont {Cleland}(2003)}]{Cleland}%
  \BibitemOpen
  \bibfield  {author} {\bibinfo {author} {\bibfnamefont {A.~N.~N.}\
  \bibnamefont {Cleland}},\ }\href@noop {} {\emph {\bibinfo {title}
  {Foundations of Nanomechanics}}}\ (\bibinfo  {publisher} {Springer, Berlin},\
  \bibinfo {year} {2003})\BibitemShut {NoStop}%
\bibitem [{\citenamefont {Kleckner}\ and\ \citenamefont
  {Bouwmeester}(2006)}]{Kleckner2006}%
  \BibitemOpen
  \bibfield  {author} {\bibinfo {author} {\bibfnamefont {D.}~\bibnamefont
  {Kleckner}}\ and\ \bibinfo {author} {\bibfnamefont {D.}~\bibnamefont
  {Bouwmeester}},\ }\href@noop {} {\bibfield  {journal} {\bibinfo  {journal}
  {Nature}\ }\textbf {\bibinfo {volume} {444}},\ \bibinfo {pages} {75}
  (\bibinfo {year} {2006})}\BibitemShut {NoStop}%
\bibitem [{\citenamefont {Poggio}\ \emph {et~al.}(2007)\citenamefont {Poggio},
  \citenamefont {Degen}, \citenamefont {Mamin},\ and\ \citenamefont
  {Rugar}}]{Poggio2007}%
  \BibitemOpen
  \bibfield  {author} {\bibinfo {author} {\bibfnamefont {M.}~\bibnamefont
  {Poggio}}, \bibinfo {author} {\bibfnamefont {C.}~\bibnamefont {Degen}},
  \bibinfo {author} {\bibfnamefont {H.}~\bibnamefont {Mamin}}, \ and\ \bibinfo
  {author} {\bibfnamefont {D.}~\bibnamefont {Rugar}},\ }\href@noop {}
  {\bibfield  {journal} {\bibinfo  {journal} {Phys. Rev. Lett.}\ }\textbf
  {\bibinfo {volume} {99}},\ \bibinfo {pages} {017201} (\bibinfo {year}
  {2007})}\BibitemShut {NoStop}%
\bibitem [{\citenamefont {Teufel}\ \emph {et~al.}(2008)\citenamefont {Teufel},
  \citenamefont {Harlow}, \citenamefont {Regal},\ and\ \citenamefont
  {Lehnert}}]{Teufel2008}%
  \BibitemOpen
  \bibfield  {author} {\bibinfo {author} {\bibfnamefont {J.}~\bibnamefont
  {Teufel}}, \bibinfo {author} {\bibfnamefont {J.}~\bibnamefont {Harlow}},
  \bibinfo {author} {\bibfnamefont {C.}~\bibnamefont {Regal}}, \ and\ \bibinfo
  {author} {\bibfnamefont {K.}~\bibnamefont {Lehnert}},\ }\href@noop {}
  {\bibfield  {journal} {\bibinfo  {journal} {Phys. Rev. Lett.}\ }\textbf
  {\bibinfo {volume} {101}},\ \bibinfo {pages} {197203} (\bibinfo {year}
  {2008})}\BibitemShut {NoStop}%
\bibitem [{\citenamefont {Arcizet}\ \emph {et~al.}(2006)\citenamefont
  {Arcizet}, \citenamefont {Cohadon}, \citenamefont {Briant}, \citenamefont
  {Pinard},\ and\ \citenamefont {Heidmann}}]{Arcizet2006}%
  \BibitemOpen
  \bibfield  {author} {\bibinfo {author} {\bibfnamefont {O.}~\bibnamefont
  {Arcizet}}, \bibinfo {author} {\bibfnamefont {P.~F.}\ \bibnamefont
  {Cohadon}}, \bibinfo {author} {\bibfnamefont {T.}~\bibnamefont {Briant}},
  \bibinfo {author} {\bibfnamefont {M.}~\bibnamefont {Pinard}}, \ and\ \bibinfo
  {author} {\bibfnamefont {A.}~\bibnamefont {Heidmann}},\ }\href@noop {}
  {\bibfield  {journal} {\bibinfo  {journal} {Nature}\ }\textbf {\bibinfo
  {volume} {444}},\ \bibinfo {pages} {71} (\bibinfo {year} {2006})}\BibitemShut
  {NoStop}%
\bibitem [{\citenamefont {Teufel}\ \emph
  {et~al.}(2011{\natexlab{a}})\citenamefont {Teufel}, \citenamefont {Donner},
  \citenamefont {Li}, \citenamefont {Harlow}, \citenamefont {Allman},
  \citenamefont {Cicak}, \citenamefont {Sirois}, \citenamefont {Whittaker},
  \citenamefont {Lehnert},\ and\ \citenamefont {Simmonds}}]{Teufel2011}%
  \BibitemOpen
  \bibfield  {author} {\bibinfo {author} {\bibfnamefont {J.~D.}\ \bibnamefont
  {Teufel}}, \bibinfo {author} {\bibfnamefont {T.}~\bibnamefont {Donner}},
  \bibinfo {author} {\bibfnamefont {D.}~\bibnamefont {Li}}, \bibinfo {author}
  {\bibfnamefont {J.~W.}\ \bibnamefont {Harlow}}, \bibinfo {author}
  {\bibfnamefont {M.~S.}\ \bibnamefont {Allman}}, \bibinfo {author}
  {\bibfnamefont {K.}~\bibnamefont {Cicak}}, \bibinfo {author} {\bibfnamefont
  {A.~J.}\ \bibnamefont {Sirois}}, \bibinfo {author} {\bibfnamefont {J.~D.}\
  \bibnamefont {Whittaker}}, \bibinfo {author} {\bibfnamefont {K.~W.}\
  \bibnamefont {Lehnert}}, \ and\ \bibinfo {author} {\bibfnamefont {R.~W.}\
  \bibnamefont {Simmonds}},\ }\href@noop {} {\bibfield  {journal} {\bibinfo
  {journal} {Nature}\ }\textbf {\bibinfo {volume} {475}},\ \bibinfo {pages}
  {359} (\bibinfo {year} {2011}{\natexlab{a}})}\BibitemShut {NoStop}%
\bibitem [{\citenamefont {Cohadon}\ \emph {et~al.}(1999)\citenamefont
  {Cohadon}, \citenamefont {Heidmann},\ and\ \citenamefont
  {Pinard}}]{PhysRevLett.83.3174}%
  \BibitemOpen
  \bibfield  {author} {\bibinfo {author} {\bibfnamefont {P.~F.}\ \bibnamefont
  {Cohadon}}, \bibinfo {author} {\bibfnamefont {A.}~\bibnamefont {Heidmann}}, \
  and\ \bibinfo {author} {\bibfnamefont {M.}~\bibnamefont {Pinard}},\
  }\href@noop {} {\bibfield  {journal} {\bibinfo  {journal} {Phys. Rev. Lett.}\
  }\textbf {\bibinfo {volume} {83}},\ \bibinfo {pages} {3174} (\bibinfo {year}
  {1999})}\BibitemShut {NoStop}%
\bibitem [{\citenamefont {Gigan}\ \emph {et~al.}(2006)\citenamefont {Gigan},
  \citenamefont {B{\"o}hm}, \citenamefont {Paternostro}, \citenamefont
  {Blaser}, \citenamefont {Langer}, \citenamefont {Hertzberg}, \citenamefont
  {Schwab}, \citenamefont {B{\"a}uerle}, \citenamefont {Aspelmeyer},\ and\
  \citenamefont {Zeilinger}}]{gigan2006self}%
  \BibitemOpen
  \bibfield  {author} {\bibinfo {author} {\bibfnamefont {S.}~\bibnamefont
  {Gigan}}, \bibinfo {author} {\bibfnamefont {H.}~\bibnamefont {B{\"o}hm}},
  \bibinfo {author} {\bibfnamefont {M.}~\bibnamefont {Paternostro}}, \bibinfo
  {author} {\bibfnamefont {F.}~\bibnamefont {Blaser}}, \bibinfo {author}
  {\bibfnamefont {G.}~\bibnamefont {Langer}}, \bibinfo {author} {\bibfnamefont
  {J.}~\bibnamefont {Hertzberg}}, \bibinfo {author} {\bibfnamefont
  {K.}~\bibnamefont {Schwab}}, \bibinfo {author} {\bibfnamefont
  {D.}~\bibnamefont {B{\"a}uerle}}, \bibinfo {author} {\bibfnamefont
  {M.}~\bibnamefont {Aspelmeyer}}, \ and\ \bibinfo {author} {\bibfnamefont
  {A.}~\bibnamefont {Zeilinger}},\ }\href@noop {} {\bibfield  {journal}
  {\bibinfo  {journal} {Nature}\ }\textbf {\bibinfo {volume} {444}},\ \bibinfo
  {pages} {67} (\bibinfo {year} {2006})}\BibitemShut {NoStop}%
\bibitem [{\citenamefont {Brown}\ \emph {et~al.}(2007)\citenamefont {Brown},
  \citenamefont {Britton}, \citenamefont {Epstein}, \citenamefont {Chiaverini},
  \citenamefont {Leibfried},\ and\ \citenamefont
  {Wineland}}]{brown2007passive}%
  \BibitemOpen
  \bibfield  {author} {\bibinfo {author} {\bibfnamefont {K.}~\bibnamefont
  {Brown}}, \bibinfo {author} {\bibfnamefont {J.}~\bibnamefont {Britton}},
  \bibinfo {author} {\bibfnamefont {R.}~\bibnamefont {Epstein}}, \bibinfo
  {author} {\bibfnamefont {J.}~\bibnamefont {Chiaverini}}, \bibinfo {author}
  {\bibfnamefont {D.}~\bibnamefont {Leibfried}}, \ and\ \bibinfo {author}
  {\bibfnamefont {D.}~\bibnamefont {Wineland}},\ }\href@noop {} {\bibfield
  {journal} {\bibinfo  {journal} {Phys. Rev. Lett.}\ }\textbf {\bibinfo
  {volume} {99}},\ \bibinfo {pages} {137205} (\bibinfo {year}
  {2007})}\BibitemShut {NoStop}%
\bibitem [{\citenamefont {Xue}\ \emph {et~al.}(2007)\citenamefont {Xue},
  \citenamefont {Wang}, \citenamefont {Liu},\ and\ \citenamefont
  {Nori}}]{xue2007cooling}%
  \BibitemOpen
  \bibfield  {author} {\bibinfo {author} {\bibfnamefont {F.}~\bibnamefont
  {Xue}}, \bibinfo {author} {\bibfnamefont {Y.}~\bibnamefont {Wang}}, \bibinfo
  {author} {\bibfnamefont {Y.~X.}\ \bibnamefont {Liu}}, \ and\ \bibinfo
  {author} {\bibfnamefont {F.}~\bibnamefont {Nori}},\ }\href@noop {} {\bibfield
   {journal} {\bibinfo  {journal} {Phys. Rev. B}\ }\textbf {\bibinfo {volume}
  {76}},\ \bibinfo {pages} {205302} (\bibinfo {year} {2007})}\BibitemShut
  {NoStop}%
\bibitem [{\citenamefont {Zhang}\ \emph {et~al.}(2009)\citenamefont {Zhang},
  \citenamefont {Liu},\ and\ \citenamefont {Nori}}]{zhang2009cooling}%
  \BibitemOpen
  \bibfield  {author} {\bibinfo {author} {\bibfnamefont {J.}~\bibnamefont
  {Zhang}}, \bibinfo {author} {\bibfnamefont {Y.}~\bibnamefont {Liu}}, \ and\
  \bibinfo {author} {\bibfnamefont {F.}~\bibnamefont {Nori}},\ }\href@noop {}
  {\bibfield  {journal} {\bibinfo  {journal} {Phys. Rev. A}\ }\textbf {\bibinfo
  {volume} {79}},\ \bibinfo {pages} {052102} (\bibinfo {year}
  {2009})}\BibitemShut {NoStop}%
\bibitem [{\citenamefont {Xiang}\ \emph {et~al.}(2013)\citenamefont {Xiang},
  \citenamefont {Ashhab}, \citenamefont {You},\ and\ \citenamefont
  {Nori}}]{Xiang2013}%
  \BibitemOpen
  \bibfield  {author} {\bibinfo {author} {\bibfnamefont {Z.~L.}\ \bibnamefont
  {Xiang}}, \bibinfo {author} {\bibfnamefont {S.}~\bibnamefont {Ashhab}},
  \bibinfo {author} {\bibfnamefont {J.~Q.}\ \bibnamefont {You}}, \ and\
  \bibinfo {author} {\bibfnamefont {F.}~\bibnamefont {Nori}},\ }\href@noop {}
  {\bibfield  {journal} {\bibinfo  {journal} {Rev. Mod. Phys.}\ }\textbf
  {\bibinfo {volume} {85}},\ \bibinfo {pages} {623} (\bibinfo {year}
  {2013})}\BibitemShut {NoStop}%
\bibitem [{\citenamefont {Hammerer}\ \emph {et~al.}(2009)\citenamefont
  {Hammerer}, \citenamefont {Wallquist}, \citenamefont {Genes}, \citenamefont
  {Ludwig}, \citenamefont {Marquardt}, \citenamefont {Treutlein}, \citenamefont
  {Zoller}, \citenamefont {Ye},\ and\ \citenamefont {Kimble}}]{Hammerer2009}%
  \BibitemOpen
  \bibfield  {author} {\bibinfo {author} {\bibfnamefont {K.}~\bibnamefont
  {Hammerer}}, \bibinfo {author} {\bibfnamefont {M.}~\bibnamefont {Wallquist}},
  \bibinfo {author} {\bibfnamefont {C.}~\bibnamefont {Genes}}, \bibinfo
  {author} {\bibfnamefont {M.}~\bibnamefont {Ludwig}}, \bibinfo {author}
  {\bibfnamefont {F.}~\bibnamefont {Marquardt}}, \bibinfo {author}
  {\bibfnamefont {P.}~\bibnamefont {Treutlein}}, \bibinfo {author}
  {\bibfnamefont {P.}~\bibnamefont {Zoller}}, \bibinfo {author} {\bibfnamefont
  {J.}~\bibnamefont {Ye}}, \ and\ \bibinfo {author} {\bibfnamefont
  {H.}~\bibnamefont {Kimble}},\ }\href@noop {} {\bibfield  {journal} {\bibinfo
  {journal} {Phys. Rev. Lett.}\ }\textbf {\bibinfo {volume} {103}},\ \bibinfo
  {pages} {063005} (\bibinfo {year} {2009})}\BibitemShut {NoStop}%
\bibitem [{\citenamefont {Wallquist}\ \emph {et~al.}(2010)\citenamefont
  {Wallquist}, \citenamefont {Hammerer}, \citenamefont {Zoller}, \citenamefont
  {Genes}, \citenamefont {Ludwig}, \citenamefont {Marquardt}, \citenamefont
  {Treutlein}, \citenamefont {Ye},\ and\ \citenamefont
  {Kimble}}]{wallquist2010single}%
  \BibitemOpen
  \bibfield  {author} {\bibinfo {author} {\bibfnamefont {M.}~\bibnamefont
  {Wallquist}}, \bibinfo {author} {\bibfnamefont {K.}~\bibnamefont {Hammerer}},
  \bibinfo {author} {\bibfnamefont {P.}~\bibnamefont {Zoller}}, \bibinfo
  {author} {\bibfnamefont {C.}~\bibnamefont {Genes}}, \bibinfo {author}
  {\bibfnamefont {M.}~\bibnamefont {Ludwig}}, \bibinfo {author} {\bibfnamefont
  {F.}~\bibnamefont {Marquardt}}, \bibinfo {author} {\bibfnamefont
  {P.}~\bibnamefont {Treutlein}}, \bibinfo {author} {\bibfnamefont
  {J.}~\bibnamefont {Ye}}, \ and\ \bibinfo {author} {\bibfnamefont
  {H.}~\bibnamefont {Kimble}},\ }\href@noop {} {\bibfield  {journal} {\bibinfo
  {journal} {Phys. Rev. A}\ }\textbf {\bibinfo {volume} {81}},\ \bibinfo
  {pages} {023816} (\bibinfo {year} {2010})}\BibitemShut {NoStop}%
\bibitem [{\citenamefont {Chang}\ \emph {et~al.}(2009)\citenamefont {Chang},
  \citenamefont {Ian},\ and\ \citenamefont {Sun}}]{chang2009triple}%
  \BibitemOpen
  \bibfield  {author} {\bibinfo {author} {\bibfnamefont {Y.}~\bibnamefont
  {Chang}}, \bibinfo {author} {\bibfnamefont {H.}~\bibnamefont {Ian}}, \ and\
  \bibinfo {author} {\bibfnamefont {C.}~\bibnamefont {Sun}},\ }\href@noop {}
  {\bibfield  {journal} {\bibinfo  {journal} {J. Phys. B: At. Mol. Opt. Phys.}\
  }\textbf {\bibinfo {volume} {42}},\ \bibinfo {pages} {215502} (\bibinfo
  {year} {2009})}\BibitemShut {NoStop}%
\bibitem [{\citenamefont {Breyer}\ and\ \citenamefont
  {Bienert}(2012)}]{breyer2012light}%
  \BibitemOpen
  \bibfield  {author} {\bibinfo {author} {\bibfnamefont {D.}~\bibnamefont
  {Breyer}}\ and\ \bibinfo {author} {\bibfnamefont {M.}~\bibnamefont
  {Bienert}},\ }\href@noop {} {\bibfield  {journal} {\bibinfo  {journal} {Phys.
  Rev. A}\ }\textbf {\bibinfo {volume} {86}},\ \bibinfo {pages} {053819}
  (\bibinfo {year} {2012})}\BibitemShut {NoStop}%
\bibitem [{\citenamefont {Barzanjeh}\ \emph {et~al.}(2011)\citenamefont
  {Barzanjeh}, \citenamefont {Naderi},\ and\ \citenamefont
  {Soltanolkotabi}}]{barzanjeh2011steady}%
  \BibitemOpen
  \bibfield  {author} {\bibinfo {author} {\bibfnamefont {S.}~\bibnamefont
  {Barzanjeh}}, \bibinfo {author} {\bibfnamefont {M.}~\bibnamefont {Naderi}}, \
  and\ \bibinfo {author} {\bibfnamefont {M.}~\bibnamefont {Soltanolkotabi}},\
  }\href@noop {} {\bibfield  {journal} {\bibinfo  {journal} {Phys. Rev. A}\
  }\textbf {\bibinfo {volume} {84}},\ \bibinfo {pages} {063850} (\bibinfo
  {year} {2011})}\BibitemShut {NoStop}%
\bibitem [{\citenamefont {Ian}\ \emph {et~al.}(2008)\citenamefont {Ian},
  \citenamefont {Gong}, \citenamefont {Liu}, \citenamefont {Sun},\ and\
  \citenamefont {Nori}}]{ian2008cavity}%
  \BibitemOpen
  \bibfield  {author} {\bibinfo {author} {\bibfnamefont {H.}~\bibnamefont
  {Ian}}, \bibinfo {author} {\bibfnamefont {Z.}~\bibnamefont {Gong}}, \bibinfo
  {author} {\bibfnamefont {Y.-x.}\ \bibnamefont {Liu}}, \bibinfo {author}
  {\bibfnamefont {C.}~\bibnamefont {Sun}}, \ and\ \bibinfo {author}
  {\bibfnamefont {F.}~\bibnamefont {Nori}},\ }\href@noop {} {\bibfield
  {journal} {\bibinfo  {journal} {Phys. Rev. A}\ }\textbf {\bibinfo {volume}
  {78}},\ \bibinfo {pages} {013824} (\bibinfo {year} {2008})}\BibitemShut
  {NoStop}%
\bibitem [{\citenamefont {Schleier-Smith}\ \emph {et~al.}(2011)\citenamefont
  {Schleier-Smith}, \citenamefont {Leroux}, \citenamefont {Zhang},
  \citenamefont {Van~Camp},\ and\ \citenamefont
  {Vuleti{\'c}}}]{schleier2011optomechanical}%
  \BibitemOpen
  \bibfield  {author} {\bibinfo {author} {\bibfnamefont {M.~H.}\ \bibnamefont
  {Schleier-Smith}}, \bibinfo {author} {\bibfnamefont {I.~D.}\ \bibnamefont
  {Leroux}}, \bibinfo {author} {\bibfnamefont {H.}~\bibnamefont {Zhang}},
  \bibinfo {author} {\bibfnamefont {M.~A.}\ \bibnamefont {Van~Camp}}, \ and\
  \bibinfo {author} {\bibfnamefont {V.}~\bibnamefont {Vuleti{\'c}}},\
  }\href@noop {} {\bibfield  {journal} {\bibinfo  {journal} {Phys. Rev. Lett.}\
  }\textbf {\bibinfo {volume} {107}},\ \bibinfo {pages} {143005} (\bibinfo
  {year} {2011})}\BibitemShut {NoStop}%
\bibitem [{\citenamefont {Genes}\ \emph
  {et~al.}(2008{\natexlab{a}})\citenamefont {Genes}, \citenamefont {Vitali},\
  and\ \citenamefont {Tombesi}}]{Genes2008}%
  \BibitemOpen
  \bibfield  {author} {\bibinfo {author} {\bibfnamefont {C.}~\bibnamefont
  {Genes}}, \bibinfo {author} {\bibfnamefont {D.}~\bibnamefont {Vitali}}, \
  and\ \bibinfo {author} {\bibfnamefont {P.}~\bibnamefont {Tombesi}},\
  }\href@noop {} {\bibfield  {journal} {\bibinfo  {journal} {Phys. Rev. A}\
  }\textbf {\bibinfo {volume} {77}},\ \bibinfo {pages} {050307} (\bibinfo
  {year} {2008}{\natexlab{a}})}\BibitemShut {NoStop}%
\bibitem [{\citenamefont {Tian}\ and\ \citenamefont {Zoller}(2004)}]{Tian2004}%
  \BibitemOpen
  \bibfield  {author} {\bibinfo {author} {\bibfnamefont {L.}~\bibnamefont
  {Tian}}\ and\ \bibinfo {author} {\bibfnamefont {P.}~\bibnamefont {Zoller}},\
  }\href@noop {} {\bibfield  {journal} {\bibinfo  {journal} {Phys. Rev. Lett.}\
  }\textbf {\bibinfo {volume} {93}} (\bibinfo {year} {2004})}\BibitemShut
  {NoStop}%
\bibitem [{\citenamefont {Bhattacharya}\ \emph {et~al.}(2010)\citenamefont
  {Bhattacharya}, \citenamefont {Singh}, \citenamefont {Giscard},\ and\
  \citenamefont {Meystre}}]{bhattacharya2010optomechanical}%
  \BibitemOpen
  \bibfield  {author} {\bibinfo {author} {\bibfnamefont {M.}~\bibnamefont
  {Bhattacharya}}, \bibinfo {author} {\bibfnamefont {S.}~\bibnamefont {Singh}},
  \bibinfo {author} {\bibfnamefont {P.-L.}\ \bibnamefont {Giscard}}, \ and\
  \bibinfo {author} {\bibfnamefont {P.}~\bibnamefont {Meystre}},\ }\href@noop
  {} {\bibfield  {journal} {\bibinfo  {journal} {Laser Phys.}\ }\textbf
  {\bibinfo {volume} {20}},\ \bibinfo {pages} {57} (\bibinfo {year}
  {2010})}\BibitemShut {NoStop}%
\bibitem [{\citenamefont {Singh}\ \emph {et~al.}(2008)\citenamefont {Singh},
  \citenamefont {Bhattacharya}, \citenamefont {Dutta},\ and\ \citenamefont
  {Meystre}}]{singh2008coupling}%
  \BibitemOpen
  \bibfield  {author} {\bibinfo {author} {\bibfnamefont {S.}~\bibnamefont
  {Singh}}, \bibinfo {author} {\bibfnamefont {M.}~\bibnamefont {Bhattacharya}},
  \bibinfo {author} {\bibfnamefont {O.}~\bibnamefont {Dutta}}, \ and\ \bibinfo
  {author} {\bibfnamefont {P.}~\bibnamefont {Meystre}},\ }\href@noop {}
  {\bibfield  {journal} {\bibinfo  {journal} {Phys. Rev. Lett.}\ }\textbf
  {\bibinfo {volume} {101}},\ \bibinfo {pages} {263603} (\bibinfo {year}
  {2008})}\BibitemShut {NoStop}%
\bibitem [{\citenamefont {Ciaramicoli}\ \emph {et~al.}(2007)\citenamefont
  {Ciaramicoli}, \citenamefont {Marzoli},\ and\ \citenamefont
  {Tombesi}}]{PhysRevA.75.032348}%
  \BibitemOpen
  \bibfield  {author} {\bibinfo {author} {\bibfnamefont {G.}~\bibnamefont
  {Ciaramicoli}}, \bibinfo {author} {\bibfnamefont {I.}~\bibnamefont
  {Marzoli}}, \ and\ \bibinfo {author} {\bibfnamefont {P.}~\bibnamefont
  {Tombesi}},\ }\href@noop {} {\bibfield  {journal} {\bibinfo  {journal} {Phys.
  Rev. A}\ }\textbf {\bibinfo {volume} {75}},\ \bibinfo {pages} {032348}
  (\bibinfo {year} {2007})}\BibitemShut {NoStop}%
\bibitem [{\citenamefont {Vitali}\ \emph {et~al.}(2007)\citenamefont {Vitali},
  \citenamefont {Tombesi}, \citenamefont {Woolley}, \citenamefont {Doherty},\
  and\ \citenamefont {Milburn}}]{PhysRevA.76.042336}%
  \BibitemOpen
  \bibfield  {author} {\bibinfo {author} {\bibfnamefont {D.}~\bibnamefont
  {Vitali}}, \bibinfo {author} {\bibfnamefont {P.}~\bibnamefont {Tombesi}},
  \bibinfo {author} {\bibfnamefont {M.~J.}\ \bibnamefont {Woolley}}, \bibinfo
  {author} {\bibfnamefont {A.~C.}\ \bibnamefont {Doherty}}, \ and\ \bibinfo
  {author} {\bibfnamefont {G.~J.}\ \bibnamefont {Milburn}},\ }\href@noop {}
  {\bibfield  {journal} {\bibinfo  {journal} {Phys. Rev. A}\ }\textbf {\bibinfo
  {volume} {76}},\ \bibinfo {pages} {042336} (\bibinfo {year}
  {2007})}\BibitemShut {NoStop}%
\bibitem [{\citenamefont {Kielpinski}\ \emph {et~al.}(2012)\citenamefont
  {Kielpinski}, \citenamefont {Kafri}, \citenamefont {Woolley}, \citenamefont
  {Milburn},\ and\ \citenamefont {Taylor}}]{kielpinski2012quantum}%
  \BibitemOpen
  \bibfield  {author} {\bibinfo {author} {\bibfnamefont {D.}~\bibnamefont
  {Kielpinski}}, \bibinfo {author} {\bibfnamefont {D.}~\bibnamefont {Kafri}},
  \bibinfo {author} {\bibfnamefont {M.}~\bibnamefont {Woolley}}, \bibinfo
  {author} {\bibfnamefont {G.}~\bibnamefont {Milburn}}, \ and\ \bibinfo
  {author} {\bibfnamefont {J.}~\bibnamefont {Taylor}},\ }\href@noop {}
  {\bibfield  {journal} {\bibinfo  {journal} {Phys. Rev. Lett.}\ }\textbf
  {\bibinfo {volume} {108}},\ \bibinfo {pages} {130504} (\bibinfo {year}
  {2012})}\BibitemShut {NoStop}%
\bibitem [{\citenamefont {Daniilidis}\ and\ \citenamefont
  {H\"{a}ffner}(2013)}]{Daniilidis2013}%
  \BibitemOpen
  \bibfield  {author} {\bibinfo {author} {\bibfnamefont {N.}~\bibnamefont
  {Daniilidis}}\ and\ \bibinfo {author} {\bibfnamefont {H.}~\bibnamefont
  {H\"{a}ffner}},\ }\href@noop {} {\bibfield  {journal} {\bibinfo  {journal}
  {Annu. Rev. Condens. Matter Phys.}\ }\textbf {\bibinfo {volume} {4}},\
  \bibinfo {pages} {83} (\bibinfo {year} {2013})}\BibitemShut {NoStop}%
\bibitem [{\citenamefont {Gangopadhyay}\ and\ \citenamefont
  {Ray}(1990)}]{Gan1990}%
  \BibitemOpen
  \bibfield  {author} {\bibinfo {author} {\bibfnamefont {G.}~\bibnamefont
  {Gangopadhyay}}\ and\ \bibinfo {author} {\bibfnamefont {D.~S.}\ \bibnamefont
  {Ray}},\ }\href@noop {} {\bibfield  {journal} {\bibinfo  {journal} {Phys.
  Rev. A}\ }\textbf {\bibinfo {volume} {41}},\ \bibinfo {pages} {6429}
  (\bibinfo {year} {1990})}\BibitemShut {NoStop}%
\bibitem [{\citenamefont {Gangopadhyay}\ and\ \citenamefont
  {Ray}(1991)}]{Gan1991}%
  \BibitemOpen
  \bibfield  {author} {\bibinfo {author} {\bibfnamefont {G.}~\bibnamefont
  {Gangopadhyay}}\ and\ \bibinfo {author} {\bibfnamefont {D.~S.}\ \bibnamefont
  {Ray}},\ }\href@noop {} {\bibfield  {journal} {\bibinfo  {journal} {Phys.
  Rev. A}\ }\textbf {\bibinfo {volume} {43}},\ \bibinfo {pages} {6424}
  (\bibinfo {year} {1991})}\BibitemShut {NoStop}%
\bibitem [{\citenamefont {Dong}\ \emph {et~al.}(2003)\citenamefont {Dong},
  \citenamefont {Tang},\ and\ \citenamefont {Sun}}]{dong2003controllability}%
  \BibitemOpen
  \bibfield  {author} {\bibinfo {author} {\bibfnamefont {S.-H.}\ \bibnamefont
  {Dong}}, \bibinfo {author} {\bibfnamefont {Y.}~\bibnamefont {Tang}}, \ and\
  \bibinfo {author} {\bibfnamefont {G.-H.}\ \bibnamefont {Sun}},\ }\href@noop
  {} {\bibfield  {journal} {\bibinfo  {journal} {Phys. Lett. A}\ }\textbf
  {\bibinfo {volume} {320}},\ \bibinfo {pages} {145} (\bibinfo {year}
  {2003})}\BibitemShut {NoStop}%
\bibitem [{\citenamefont {Angelova}\ and\ \citenamefont
  {Hussin}(2008)}]{angelova2008generalized}%
  \BibitemOpen
  \bibfield  {author} {\bibinfo {author} {\bibfnamefont {M.}~\bibnamefont
  {Angelova}}\ and\ \bibinfo {author} {\bibfnamefont {V.}~\bibnamefont
  {Hussin}},\ }\href@noop {} {\bibfield  {journal} {\bibinfo  {journal} {J.
  Phys. A: Math. Theor.}\ }\textbf {\bibinfo {volume} {41}},\ \bibinfo {pages}
  {304016} (\bibinfo {year} {2008})}\BibitemShut {NoStop}%
\bibitem [{\citenamefont {Noggle}(1996)}]{Noggle1996}%
  \BibitemOpen
  \bibfield  {author} {\bibinfo {author} {\bibfnamefont {J.~H.}\ \bibnamefont
  {Noggle}},\ }\href@noop {} {\emph {\bibinfo {title} {Physical Chemistry, 3rd
  Edition}}}\ (\bibinfo  {publisher} {New York},\ \bibinfo {year}
  {1996})\BibitemShut {NoStop}%
\bibitem [{\citenamefont {Schleich}(2001)}]{schleich2011quantum}%
  \BibitemOpen
  \bibfield  {author} {\bibinfo {author} {\bibfnamefont {W.~P.}\ \bibnamefont
  {Schleich}},\ }\href@noop {} {\emph {\bibinfo {title} {Quantum optics in
  phase space}}}\ (\bibinfo  {publisher} {WILEY-VCH},\ \bibinfo {year}
  {2001})\BibitemShut {NoStop}%
\bibitem [{\citenamefont {Gardiner}\ and\ \citenamefont
  {Zoller}(2004)}]{gardiner2004quantum}%
  \BibitemOpen
  \bibfield  {author} {\bibinfo {author} {\bibfnamefont {C.}~\bibnamefont
  {Gardiner}}\ and\ \bibinfo {author} {\bibfnamefont {P.}~\bibnamefont
  {Zoller}},\ }\href@noop {} {\emph {\bibinfo {title} {Quantum noise: a
  handbook of Markovian and non-Markovian quantum stochastic methods with
  applications to quantum optics}}},\ Vol.~\bibinfo {volume} {56}\ (\bibinfo
  {publisher} {Springer},\ \bibinfo {year} {2004})\BibitemShut {NoStop}%
\bibitem [{\citenamefont {Holmes}\ and\ \citenamefont
  {Milburn}(2009)}]{holmes2009parametric}%
  \BibitemOpen
  \bibfield  {author} {\bibinfo {author} {\bibfnamefont {C.}~\bibnamefont
  {Holmes}}\ and\ \bibinfo {author} {\bibfnamefont {G.}~\bibnamefont
  {Milburn}},\ }\href@noop {} {\bibfield  {journal} {\bibinfo  {journal}
  {Fortschritte der Physik}\ }\textbf {\bibinfo {volume} {57}},\ \bibinfo
  {pages} {1052} (\bibinfo {year} {2009})}\BibitemShut {NoStop}%
\bibitem [{\citenamefont {Genes}\ \emph
  {et~al.}(2008{\natexlab{b}})\citenamefont {Genes}, \citenamefont {Vitali},
  \citenamefont {Tombesi}, \citenamefont {Gigan},\ and\ \citenamefont
  {Aspelmeyer}}]{genes2008ground}%
  \BibitemOpen
  \bibfield  {author} {\bibinfo {author} {\bibfnamefont {C.}~\bibnamefont
  {Genes}}, \bibinfo {author} {\bibfnamefont {D.}~\bibnamefont {Vitali}},
  \bibinfo {author} {\bibfnamefont {P.}~\bibnamefont {Tombesi}}, \bibinfo
  {author} {\bibfnamefont {S.}~\bibnamefont {Gigan}}, \ and\ \bibinfo {author}
  {\bibfnamefont {M.}~\bibnamefont {Aspelmeyer}},\ }\href@noop {} {\bibfield
  {journal} {\bibinfo  {journal} {Phys. Rev. A}\ }\textbf {\bibinfo {volume}
  {77}},\ \bibinfo {pages} {033804} (\bibinfo {year}
  {2008}{\natexlab{b}})}\BibitemShut {NoStop}%
\bibitem [{\citenamefont {Teufel}\ \emph
  {et~al.}(2011{\natexlab{b}})\citenamefont {Teufel}, \citenamefont {Donner},
  \citenamefont {Li}, \citenamefont {Harlow}, \citenamefont {Allman},
  \citenamefont {Cicak}, \citenamefont {Sirois}, \citenamefont {Whittaker},
  \citenamefont {Lehnert},\ and\ \citenamefont {Simmonds}}]{teufel}%
  \BibitemOpen
  \bibfield  {author} {\bibinfo {author} {\bibfnamefont {J.}~\bibnamefont
  {Teufel}}, \bibinfo {author} {\bibfnamefont {T.}~\bibnamefont {Donner}},
  \bibinfo {author} {\bibfnamefont {D.}~\bibnamefont {Li}}, \bibinfo {author}
  {\bibfnamefont {J.}~\bibnamefont {Harlow}}, \bibinfo {author} {\bibfnamefont
  {M.}~\bibnamefont {Allman}}, \bibinfo {author} {\bibfnamefont
  {K.}~\bibnamefont {Cicak}}, \bibinfo {author} {\bibfnamefont
  {A.}~\bibnamefont {Sirois}}, \bibinfo {author} {\bibfnamefont
  {J.}~\bibnamefont {Whittaker}}, \bibinfo {author} {\bibfnamefont
  {K.}~\bibnamefont {Lehnert}}, \ and\ \bibinfo {author} {\bibfnamefont
  {R.}~\bibnamefont {Simmonds}},\ }\href@noop {} {\bibfield  {journal}
  {\bibinfo  {journal} {Nature}\ }\textbf {\bibinfo {volume} {475}},\ \bibinfo
  {pages} {359} (\bibinfo {year} {2011}{\natexlab{b}})}\BibitemShut {NoStop}%
\bibitem [{\citenamefont {Rabl}\ \emph {et~al.}(2006)\citenamefont {Rabl},
  \citenamefont {DeMille}, \citenamefont {Doyle}, \citenamefont {Lukin},
  \citenamefont {Schoelkopf},\ and\ \citenamefont {Zoller}}]{rabl2006hybrid}%
  \BibitemOpen
  \bibfield  {author} {\bibinfo {author} {\bibfnamefont {P.}~\bibnamefont
  {Rabl}}, \bibinfo {author} {\bibfnamefont {D.}~\bibnamefont {DeMille}},
  \bibinfo {author} {\bibfnamefont {J.}~\bibnamefont {Doyle}}, \bibinfo
  {author} {\bibfnamefont {M.}~\bibnamefont {Lukin}}, \bibinfo {author}
  {\bibfnamefont {R.}~\bibnamefont {Schoelkopf}}, \ and\ \bibinfo {author}
  {\bibfnamefont {P.}~\bibnamefont {Zoller}},\ }\href@noop {} {\bibfield
  {journal} {\bibinfo  {journal} {Phys. Rev. Lett}\ }\textbf {\bibinfo {volume}
  {97}},\ \bibinfo {pages} {033003} (\bibinfo {year} {2006})}\BibitemShut
  {NoStop}%
\bibitem [{\citenamefont {Gradshteyn}\ and\ \citenamefont
  {Ryzhik}()}]{gradshteyn1980table}%
  \BibitemOpen
  \bibfield  {author} {\bibinfo {author} {\bibfnamefont {I.}~\bibnamefont
  {Gradshteyn}}\ and\ \bibinfo {author} {\bibfnamefont {I.}~\bibnamefont
  {Ryzhik}},\ }\href@noop {} {\emph {\bibinfo {title} {Table of Integrals,
  Series and Products}}}\ (\bibinfo  {publisher} {Academic, Orlando,
  1980})\BibitemShut {NoStop}%
\bibitem [{\citenamefont {Zdravkovi{\'c}}\ and\ \citenamefont
  {Satari{\'c}}(2012)}]{zdravkovic2012morse}%
  \BibitemOpen
  \bibfield  {author} {\bibinfo {author} {\bibfnamefont {S.}~\bibnamefont
  {Zdravkovi{\'c}}}\ and\ \bibinfo {author} {\bibfnamefont {M.~V.}\
  \bibnamefont {Satari{\'c}}},\ }\href@noop {} {\bibfield  {journal} {\bibinfo
  {journal} {J. Biosci}\ }\textbf {\bibinfo {volume} {37}},\ \bibinfo {pages}
  {613} (\bibinfo {year} {2012})}\BibitemShut {NoStop}%
\bibitem [{\citenamefont {Peyrard}\ and\ \citenamefont
  {Bishop}(1989)}]{peyrard1989statistical}%
  \BibitemOpen
  \bibfield  {author} {\bibinfo {author} {\bibfnamefont {M.}~\bibnamefont
  {Peyrard}}\ and\ \bibinfo {author} {\bibfnamefont {A.}~\bibnamefont
  {Bishop}},\ }\href@noop {} {\bibfield  {journal} {\bibinfo  {journal} {Phys.
  Rev. Lett.}\ }\textbf {\bibinfo {volume} {62}} (\bibinfo {year}
  {1989})}\BibitemShut {NoStop}%
\bibitem [{\citenamefont {Dauxois}(1991)}]{dauxois1991dynamics}%
  \BibitemOpen
  \bibfield  {author} {\bibinfo {author} {\bibfnamefont {T.}~\bibnamefont
  {Dauxois}},\ }\href@noop {} {\bibfield  {journal} {\bibinfo  {journal} {Phys.
  Rev. Lett.}\ }\textbf {\bibinfo {volume} {159}},\ \bibinfo {pages} {390}
  (\bibinfo {year} {1991})}\BibitemShut {NoStop}%
\end{thebibliography}%
 \bibliographystyle{apsrev4-1} 
\end{document}